\documentclass[reqno,11pt]{amsart}
\usepackage{hyperref}
\usepackage[mathscr]{eucal}
\usepackage{enumerate}
\numberwithin{equation}{section}

\newtheorem{Def}{Definition}[section]
\newtheorem{Thm}[Def]{Theorem}
\newtheorem{Prop}[Def]{Proposition}
\newtheorem{Lemma}[Def]{Lemma}

\newtheorem{Corollary}[Def]{Corollary}
\newtheorem{Example}[Def]{Example}

\newcommand{\beq}{\begin{equation}}
\newcommand{\beqn}{\begin{equation}\nonumber}
\newcommand{\eeq}{\end{equation}}
\newcommand{\Proof}{\begin{proof}}
\newcommand{\QED}{\end{proof} \noindent}

\newcommand{\R}{\mathbb{R}}
\newcommand{\Z}{\mathbb{Z}}
\newcommand{\N}{\mathcal{N}}
\newcommand{\M}{\mathcal{M}}
\newcommand{\U}{\mathcal{U}}

\newcommand{\ew}{\Big\rvert_w \hspace{-.2cm}}
\newcommand{\ewl}{\Big\rvert_{w_l} \hspace{-.4cm}}
\newcommand{\nn}{\nonumber}

\allowdisplaybreaks[1]

\title[Constrained Systems of Conservation Laws]{Constrained Systems of Conservation Laws:\\ A Geometric Theory}

\author[M.\ Reintjes]{Moritz Reintjes}
\thanks{M.R. is currently supported by FCT/Portugal through (GPSEinstein) PTDC/MAT-ANA/1275/2014 and UID/MAT/04459/2013. The mayor part of this work was done while M.R. was Post-Doctorate at IMPA in Rio de Janeiro, funded through CAPES-Brazil.}
\address{Departamento de Matem{\'a}tica \\ Instituto Superior T{\'e}cnico \\ 1049-001 Lisbon \\ Portugal}
\email{moritzreintjes@gmail.com}

\begin{document}

\begin{abstract}
We address the Riemann and Cauchy problems for systems of $n$ conservation laws in $m$ unknowns which are subject to $m-n$ constraints ($m\geq n$). Such constrained systems generalize systems of conservation laws in standard form to include various examples of conservation laws in Physics and Engineering beyond gas dynamics, e.g., multi-phase flow in porous media. We prove local well-posedness of the Riemann problem and global existence of the Cauchy problem for initial data with sufficiently small total variation, in one spatial dimension. The key to our existence theory is to generalize the $m\times n$ systems of constrained conservation laws to $n\times n$ systems of conservation laws with states taking values in an $n$-dimensional manifold and to extend Lax's theory for local existence as well as Glimm's random choice method to our geometric framework. Our resulting existence theory allows for the accumulation function to be \emph{non-invertible} across hypersurfaces. 
\end{abstract}

\maketitle
\tableofcontents

\section{Introduction}

In this paper we develop a geometric framework for proving well-posedness of the Riemann and Cauchy problems for systems of conservation laws by extending the (flat) space of states to a \emph{manifold of states}. Our motivation comes from constrained conservation laws that appear in Physics and Engineering, systems of the form
\begin{eqnarray} 
G(u)_t + F(u)_x =0, \label{constrained_system_eqn} \\
C(u)=0 , \label{constrained_system_constrained}
\end{eqnarray}
with states $u(x,t) \in \hat{\U} \subset \R^m$, $t\geq 0$, $x\in \R$, and where $G,\,F\,: \hat{\U} \rightarrow \R^n$ and $C: \hat{\U} \rightarrow \R^{m-n}$ are smooth functions and $m\geq n$. The function $C$ gives the constraints on the states and is determined by physical and chemical principles in applications. Often $C$ is not given explicitly but must be derived from the equations. In this case, until the constraints are identified, the system appears to have fewer eigenvalues than equations. The Euler equations of gas dynamics are a constrained system for the gas equation of state being a (trivial) constraint. Non-trivial examples of conservation laws in Physics or Engineering of form \eqref{constrained_system_eqn} - \eqref{constrained_system_constrained} often reflect transport of components that may coexist in liquid or gaseous form under thermodynamic equilibrium, see for instance \cite{LambertMarchesin2,LambertMarchesinBruining,MailybMarchesinBruining,LambertMarchesin}.\footnote{The discussion leading to equation (2.3) in \cite{LambertMarchesin2} nicely illustrates how constraints arise.} More generally, hyperbolic conservation laws with relaxation terms give rise to constraints, determining the equilibrium states of the relaxation system \cite{Liu2} (see also \cite{Bianchini,JinXin}).

For nonlinear systems of conservation laws shock waves form generically and it is most natural to study the Riemann problem. That is, piece-wise constant initial data containing one discontinuity.  In \cite{Lax}, Lax proved local existence for the Riemann problem for hyperbolic systems of conservation laws in \emph{standard form}, 
\beq\label{conservationlaw_standard}
u_t+f(u)_x=0.
\eeq 
Building on Lax's work, Glimm proved global existence of solutions of the Cauchy problem for \eqref{conservationlaw_standard}. 

From a theoretical perspective, assuming $G$ and $C$ are invertible, we can easily prove existence of solutions for \eqref{constrained_system_eqn} - \eqref{constrained_system_constrained}. Namely, we first solve the constraint locally, say, using the implicit function theorem which gives us a local parameterization of the solution space $C^{-1}(0)$. Secondly, we introduce the change of variable $U\equiv G(u) \in \R^n$ for all states $u$ lying in the image of the paarmeterization, which then allows us to write \eqref{constrained_system_eqn} in standard form \eqref{conservationlaw_standard} for
\beq \label{change_variables}
f(U)\equiv F\circ G^{-1}(U).
\eeq
We can now apply Lax's and Glimm's method to conclude with the existence of solutions for the Riemann and the Cauchy problem.

However, this approach has three major drawbacks. First, there exists balance laws in Petroleum Engineering for which $G$ fails to be invertible on points or hypersurfaces in state space \cite[Chapter 5.1]{KokubunMailybaev}. Secondly, in practice, inverting the accumulation function $G$ \emph{explicitly} is often an obstacle too hard to overcome, regardless whether $G^{-1}$ exists theoretically.\footnote{The Euler equations can be written in standard form, after changing variables,  however, this change of variables is non-trivial and one cannot expect to explicitly write all conservation laws of Physics or Engineering in standard form.} Thirdly, a single parameterization does generally not cover the entire solution space $C^{-1}(0)$ which could restrict the set of states connectable by shock and rarefaction curves artificially. To cover $C^{-1}(0)$ entirely, a suitable collection of such parameterizations is needed, which then defines a manifold. 

Motivated to overcome these drawbacks, our approach here is to extend Lax's and Glimm's existence theories beyond standard form \eqref{conservationlaw_standard}, instead of reducing \eqref{constrained_system_eqn}  - \eqref{constrained_system_constrained} to standard form. We thus avoid inverting the accumulation function $G$ entirely. Without introducing the variable $U\equiv G(u)$, one is automatically forced to consider the manifold of constraints as state space, unless one is willing to accept artificial restrictions on state space and dependence on the choice of a local parameterization. We therefore extend Lax's and Glimm's methods to the geometric framework of Riemannian manifolds of states, so that the resulting existence theory is independent of local parameterizations. Remarkably, it is possible to extend Lax's and Glimm's methods even further and allow for the accumulation function $G$ to be \emph{non-invertible} on a finite union of hypersurfaces in state space. In Proposition \eqref{Thm_parameterization}, we show that inverting $C$ explicitly can be circumvented in practice.

Let us finally remark that the fruitful geometric treatment of constraints on state space in Classical Mechanics has been further motivation for us to extend Lax's and Glimm's methods to a similar geometric setting. To give a broader context, Differential Geometry plays a central role in the theory of shock waves in General Relativity where (in contrast to this paper) the conservation laws of fluid dynamics are defined \emph{on} a manifold \cite{LeFloch,GroahTemple}, for instance, the question whether shock waves create essential spacetime singularities \cite{ReintjesTemple1,ReintjesTemple2,Reintjes} has a deeper geometric structure \cite{ReintjesTemple3}.
\vspace{.2cm}

We now present our main results. In order to obtain an existence theory for \eqref{constrained_system_eqn} - \eqref{constrained_system_constrained}, we consider the following $n\times n$ systems of conservation laws with states assuming values in some \emph{abstract} differentiable manifold $\M$:
\beq \label{system_cons_laws_manifold}
g(w)_t + f(w)_x =0, 
\eeq
where $w(x,t) \in \M$, and $g$ and $f$ are $C^3$ functions from $\M$ to $\R^n$, (c.f. \cite{doCamo,Spivak} for an introduction into Riemannian Geometry). The system of constrained conservation laws \eqref{constrained_system_eqn} - \eqref{constrained_system_constrained} assumes the above form by setting $u=w$ whenever $u$ lies in the manifold $C^{-1}(0)$, and by defining $g$ and $f$ to be the restrictions of $G$ and $F$ to $C^{-1}(0)$. The initial data for the Riemann problem for \eqref{system_cons_laws_manifold} is
\beq \label{Riemann_problem}
w(x,0) = \begin{cases}  w_l \,  , \ \ \ x<0 \cr  w_r \, , \ \ \  x>0 \end{cases},
\eeq
for $w_l$ and $w_r$ points in $\M$, so-called constant states. Our first theorem generalizes Lax's existence theory to our framework of a manifold of states.  

\begin{Thm} \label{Thm_manifold}
Assume $\M$ is a $C^3$ manifold and let $w_l \in \M$. Assume \eqref{system_cons_laws_manifold} is strictly hyperbolic in the sense of Definition \ref{def_hyperbolicity} and assume that each characteristic field is either genuinely nonlinear 
or linearly degenerate. Then there exists a neighborhood $\U$ of $w_l$ in $\M$ such that if $w_r\in \U$, the Riemann problem \eqref{system_cons_laws_manifold} with initial data \eqref{Riemann_problem} has a solution which consists of (at most) $n+1$ constant states in $\M$ separated by shocks, rarefaction waves or contact discontinuities, such that the shocks satisfy the Lax-admissibility condition \eqref{Lax_conditions} in $\U$. This solution is unique within the class of contact discontinuities, admissible shocks and rarefaction waves separated by (at most) $n+1$ constant states.
\end{Thm}

Our definition of strict hyperbolicity is general enough to allow for $dg$ to have a non-trivial null-space on set of measure zero, that is, invertibility of $dg$ is allowed to fail at lower dimensional surfaces, as explained in detail below Definition \ref{def_hyperbolicity}. The extended Lax method for proving Theorem \ref{Thm_manifold} is independent of local parameterizations and the construction of shock and rarefaction curves is \emph{not} restricted to coordinate neighborhoods. 
Remarkably, parameterizing the wave curves by arc-length, the most natural parameterization for curves in Riemannian Geometry, implies the $C^2$ contact between shock and rarefaction curves, c.f. Lemma \ref{shock-param-lemma}.  

Building on Theorem \ref{Thm_manifold}, we address  in Section \ref{sec_Glimm} the \emph{Cauchy problem} for \eqref{system_cons_laws_manifold} with initial data of sufficiently small total variation and prove existence of weak solutions by extending Glimm's famous random choice method to our framework, as recorded in the next theorem. We define the total variation ($T.V.(\cdot)$) and our analog of the $L^\infty$-norm ($d_\infty(\cdot,\cdot)$) in terms of the canonical distance function on $\M$, c.f. \eqref{distancefunction} - \eqref{supnorm}.

\begin{Thm} \label{Glimm_Thm_Intro}
Assume the system \eqref{system_cons_laws_manifold} is strictly hyperbolic in the sense of Definition \ref{def_hyperbolicity} and each characteristic field is genuinely non-linear or linearly degenerate in some neighborhood of some point $\bar{w}\in\M$. Given some curve                
\beq \nn
w_0: \R \rightarrow \M
\eeq 
such that $T.V.(w_0)$ and $d_\infty(w_0,\bar{w})$ are sufficiently small, then there exists a weak solution $w(x,t)\in\M$ of \eqref{system_cons_laws_manifold} for all $x\in \R$ and all $t\geq 0$ with initial data $w_0$, and there exists a constant $C>0$ such that     
\begin{eqnarray}\nn
T.V.\big(w(\cdot,t)\big) + d_\infty\big(w(\cdot,t),\bar{w}\big) &\leq & C\,\big( d_\infty(w_0,\bar{w}) + T.V.(w_0)  \big), \hspace{.3cm} \forall \ t\geq 0,  \cr
\int_{-\infty}^\infty d_\M\big( w(x,t_2),w(x,t_1) \big) dx &\leq & C\, |t_2-t_1| \ T.V.(w_0).
\end{eqnarray} 
\end{Thm}

Glimm's estimates for wave interaction extend quite naturally to our framework, since wave strength is measured by the parameters of wave curves.  We thus obtain a uniform bound on the Glimm functionals over the approximate solutions generated by the random choice method, by using Glimm's original argument modified to our framework. However, to conclude the existence of a convergent subsequence of the approximate solutions constructed with Glimm's scheme, one faces the difficulty of defining appropriate norms.  In particular, a supremums norm on the states themselves does not seem to make sense and one has to abandon convergence in $L^1$ of the approximate solutions.  We overcome these difficulties by using the canonical distance function of a Riemannian metric on $\M$ to measure the distance of states, which does not define a norm, but which allows us to define an adapted expression of total variation for curves in $\M$ and for the supremums norm on states. After extending Helly's Theorem to our framework, we obtain point-wise convergence of the approximate solutions and convergence in $L^1$ of the flux $f(w)$ and the accumulation $g(w)$, (abandoning $L^1$ convergence of states all together), which suffices to prove the existence of a weak solution to \eqref{system_cons_laws_manifold}. \vspace{.2cm}

From Theorem \ref{Thm_manifold} we obtain the following theorem regarding the constrained system \eqref{constrained_system_eqn} - \eqref{constrained_system_constrained}.

\begin{Thm} \label{Thm_constrained-system}
Let $G,F\in C^3(\hat{\U},\R^n)$ and $C\in C^3(\hat{\U},\R^{m-n})$ for some open set $\hat{\U} \subset \R^m$. Assume that for all $u \in \hat{\U}$ with $C(u)=0$         
\beq \label{Thm_constrained-system_cond1}
\text{det}\,\left[\begin{array}{c}DG(u) \cr DC(u)\end{array} \right] \neq 0, 
\eeq
($D$ denotes differentiation in $\hat{\mathcal{U}}\subset \R^m$), and such that there exists $r_k(u) \in \R^m$ and $\lambda_k(u) \in \R$, for all $k\in \{1,...,n\}$, solving 
\begin{eqnarray} 
\left( \lambda_k\, DG\, -\, DF \right) r_k &=& 0 , \label{Thm_constrained-system_cond2} \\
DC\, r_k &=& 0 \label{Thm_constrained-system_cond3},
\end{eqnarray}
with $\lambda_1(u) < ... < \lambda_n(u)$. Assume each field is genuinely non-linear or linearly degenerate. Let $u_l \in \hat{\U}$ with $C(u_l)=0$. Then, there exists a neighborhood $\U \subset \hat{\U}$ of $u_l$ such that if $u_r\in \U$ and $C(u_r)=0$, then the Riemann problem for $(u_l,u_r)$ of \eqref{constrained_system_constrained} has the unique solution stated in Theorem \ref{Thm_manifold} for the manifold $\M\equiv C^{-1}(0)$. Moreover, this existence result still holds when the assumption \eqref{Thm_constrained-system_cond1} is weakened such that $DG$ is not invertible, as specified in Definition \ref{def_hyperbolicity}.     
\end{Thm}

Theorem \ref{Thm_constrained-system} follows from Theorem \ref{Thm_manifold}, since \eqref{Thm_constrained-system_cond1} implies that $C^{-1}(0)$ defines a manifold whenever $DC$ has full rank, c.f. \cite[Theorem 5.1]{Spivak}. Equations \eqref{Thm_constrained-system_cond2} and \eqref{Thm_constrained-system_cond3} are the generalized eigenvalue problem and \eqref{Thm_constrained-system_cond3} is a consistency condition which ensures that the wave curves of the system remain in the solution space of the constraint. One can weaken condition \eqref{Thm_constrained-system_cond1} for $DG$ to be non-invertible on sets of measure zero in $\M\equiv C^{-1}(0)$ as specified by our notion of hyperbolicity in Definition \ref{def_hyperbolicity}.

The extended Glimm's method of Theorem \ref{Glimm_Thm_Intro} yields existence of weak solutions to the Cauchy problem for \eqref{constrained_system_eqn}~-~\eqref{constrained_system_constrained},  under the assumption \eqref{Thm_constrained-system_cond1}~-~\eqref{Thm_constrained-system_cond3} and corresponding small data. Since this is rather straightforward, we are content not to summarize the application of Theorem \ref{Glimm_Thm_Intro} as a separate theorem.

One can avoid inverting the constraint $C$ to obtain the solution space $C^{-1}(0)$, explicitly, since the integral curves of the $r_k$'s describe $C^{-1}(0)$ locally, as recorded in the next proposition. This follows from the consistency condition \eqref{Thm_constrained-system_cond3} together with the construction underlying Theorem \ref{Thm_constrained-system}. The result of the next proposition is very useful for applications, since it reduces the inversion problem for $C$ to methods from Linear Algebra.            

\begin{Prop} \label{Thm_parameterization}
Assume \eqref{Thm_constrained-system_cond1} - \eqref{Thm_constrained-system_cond3} hold for all $u$ in some open subset of $\R^m$.  Then, locally, $C^{-1}(0)$ equals the image of the integral curves of the eigenvectors $r_k$. Moreover, the integral curves of $r_k$ define a local parameterization of $C^{-1}(0)$ if and only if there exists a coordinate system of Riemann invariants.
\end{Prop}

We develop the geometric framework underlying Theorem \ref{Thm_manifold} in Section \ref{sec_frame} and prove Theorem \ref{Thm_manifold} in Section \ref{sec_Thm_mfd}. In Section \ref{sec_Thm_constraint}, we prove Theorem \ref{Thm_constrained-system} and Proposition \ref{Thm_parameterization} . In Section \ref{sec_Glimm}, we generalize Glimm's random choice method to our framework and prove global existence of solutions for the Cauchy problem with initial data of small total variation, recorded in Theorem \ref{Glimm_Thm}. 
In Appendix \ref{sec_preliminaries}, we give an overview of Riemannian geometry tailored to the background required for our framework.

\section{The Geometric Framework} \label{sec_frame}

\subsection{Basic Notions}

We now introduce the geometric framework underlying Theorem \ref{Thm_manifold}, first extending the notion of hyperbolicity, genuine nonlinearity and linear degeneracy to states taking values in a manifold $\M$. To begin, write \eqref{system_cons_laws_manifold} in the equivalent form         
\beq \label{system_cons_laws_manifold_2}
dg\ew (w_t) + df\ew (w_x) =0,
\eeq
where $dg$ and $df$ denote the differential of $g$ and $f$ respectively. That is, $dg$ and $df$ are point-wise linear mappings from $T_w\M$, the space of tangent vectors at $w$, to $\R^n$, c.f. \eqref{differential}. Note that $w_t,\,w_x\,\in T_w\M$, since $t \mapsto w(x,t)$ for $x$ fixed and $x\mapsto w(x,t)$ for $t$ fixed define curves on $\M$, and since the derivative of a curve with respect to its parameter is a tangent vector, c.f. \eqref{tangent_curve}. System \eqref{system_cons_laws_manifold_2} now gives rise to the eigenvalue problem
\beq \label{eigenvalue-problem_right_mfd}
\lambda_k(w)\  dg\ew \big(r_k(w)\big) = df\Big\rvert_w \hspace{-.3cm} \big(r_k(w)\big),
\eeq
where $\lambda_k(w)$ is a real number and $r_k(w) \in T_w\M$ is a vector, the so-called right-eigenvector. Thus, $\lambda_k : \M \rightarrow \R$ is a real-valued scalar and $r_k$ is a vector field on $\M$. Note that a solution $(\lambda_k, r_k)$ is independent of the choice of coordinate, which is fundamental to our framework. We mostly refer to the pair $\big(\lambda_k,r_k\big)$, but sometimes to $r_k$ alone, as the $k$-th characteristic field.

We now define the notion of hyperbolicity in our framework and thereby specify in what sense $dg|_w$ is allowed to have a non-trivial null-space. If $dg(r_k)=0$ at $w$, then, for the eigenvalue problem \eqref{eigenvalue-problem_right_mfd} to hold, $r_k$ must also lie in the null-space of $df|_w$, so that $\lambda_k(w)$ would be left undetermined at the point $w$ in $\M$. Nevertheless, if $\Sigma$, the set of points $w\in \M$ at which $dg|_w$ fails to be invertible, is a hypersurface in $\M$ and if $\lambda_k$ is continuous across $\Sigma$, then we define $\lambda_k$ on $\Sigma$ uniquely as the continuous extension of $\lambda_k$ to  $ \M$. This motivates the next definition.

\begin{Def} \label{def_hyperbolicity}
We call \eqref{system_cons_laws_manifold} a hyperbolic system of conservation laws in $\M$, if the following holds:
\begin{enumerate}[(i)]
\item  $dg|_w$ is invertible for all $w\in\M \setminus \Sigma$, where $\Sigma \subset \M$ denote the union of a finite family of co-dimension one surfaces; 
\item for each $w \in \M\setminus \Sigma$ and $k \in \{1,...,n\}$ there exists $r_k(w) \in T_w\M$ and $\lambda_k(w)\in \R$ that solve the eigenvalue problem \eqref{eigenvalue-problem_right_mfd}, such that the eigenvectors $r_k(w)$ are linearly independent;
\item $\lambda_k$ and $r_k$ can be extended as $C^2$ functions to $\M$, such that the resulting $r_k$ are linearly independent on $\M$. 
\end{enumerate}
We call \eqref{system_cons_laws_manifold} strictly hyperbolic in $\M$ if in addition, for all $w\in \M$, 
\beq \label{strict_hyperbolicity}
\lambda_1(w) < ... < \lambda_n(w).
\eeq
\end{Def}            

The essential point, regarding our assumption for $\Sigma$ in Definition \eqref{def_hyperbolicity}, is $\Sigma$ being of measure zero so that the extension of $\lambda_k$ is unique.\footnote{If $\Sigma$ were an open set, then the eigenvalue problem \eqref{eigenvalue-problem_right_mfd} should rather be considered as additional \emph{constraints} of the system and not as part of the evolution.}  
Defining $\lambda_k$ as the continuous extension then suffices to construct solutions of \eqref{system_cons_laws_manifold} following Lax's method. However, to prove the existence of shock curves we need to assume that $\Sigma$ consists of only finitely many hypersurfaces for the characteristic polynomial associated with the Hugoniot locus to have the required stability property. (Our methods also work for surfaces of higher co-dimension.) Below we give some examples of hyperbolic systems of conservation laws, in the sense of Definition \ref{def_hyperbolicity}, for which $dg$ fails to be invertible everywhere.

Whether the $k$-th characteristic field corresponds to a solution that is either a shock or a rarefaction wave, or whether it corresponds to a contact discontinuity depends on the fields being genuinely nonlinear or linearly degenerate, respectively. We now define these notions.

\begin{Def} \label{def_genuinely-nonlinear}
We say the $k$-th characteristic field is genuinely nonlinear in $\M$, if for all $w \in \M$
\beq \label{genuinely-nonlinear}
 r_k\ew \left(\lambda_k \right) > 0.
\eeq 
\end{Def}

In Definition \ref{def_genuinely-nonlinear} we use the interpretation of vectors as directional derivatives of scalar functions. That is, expressed in coordinates $y^\mu$ on a subset of $\M$, we have $r_k|_w = r^\mu_k(y(w)) \frac{\partial}{\partial y^\mu}$ and \eqref{genuinely-nonlinear} thus becomes 
\beq\nonumber
 r_k\ew \left(\lambda_k \right) = r^{\ \mu}_k\big(y(w)\big) \, \frac{\partial \left(\lambda\circ y^{-1}\right)\big( y(w) \big)}{\partial y^\mu},
\eeq
which is indeed a real number independent of the choice of coordinates $y^\mu$. In the special case $\M = \R^n$, \eqref{genuinely-nonlinear} reduces to the standard definition of genuine nonlinearity in $\R^n$, namely, 
$$
\langle \nabla \lambda_k , r_k \rangle_{\R^n} > 0.
$$

Note that \eqref{genuinely-nonlinear} leaves the freedom to scale the length of $r_k$ and this is often used to set $r_k\ew \left(\lambda_k \right) = 1$. However, we are \emph{not} using this normalization, but work with the normalization condition 
$$\langle r_k,r_k\rangle_\M=1,$$ 
in terms of the metric tensor $\langle \cdot,\cdot\rangle_\M$, which allows us to parameterize the shock and rarefaction curves with respect to arc-length. Analogously to \eqref{genuinely-nonlinear}, the following definition generalizes the notion of linear degeneracy to states in a manifold.

\begin{Def} \label{def_linearly-degenerate}
We say the $k$-th characteristic field is linearly degenerate in $\M$, if for all $w \in \M$
\beq \label{linearly degenerate}
 r_k\ew \left(\lambda_k \right) = 0.
\eeq 
\end{Def}

We assume in this paper that the characteristic fields are either linearly degenerate or genuinely non-linear to prove well-posedness of the Riemann problem and develop the framework in a simplest setting. We expect however that Liu's construction \cite{Liu} extends to our geometric framework, in particular to our notion of hyperbolicity in Definition \ref{def_hyperbolicity}, so that genuine non-linearity is allowed to fail across hypersurfaces in state space. 

To close this section, we give two examples of systems of conservation laws hyperbolic in the sense of Definition \ref{def_hyperbolicity}, for which $dg$ is not invertible.

\begin{Example} \label{example1}
\normalfont Take as accumulation and flux function in \eqref{system_cons_laws_manifold}
\beq \nn
g(u,v) \equiv \left(\begin{array}{c} \frac{1}{2}u^2 \cr \frac{1}{3}v^3 \end{array} \right)  
\hspace{1cm} \text{and} \hspace{1cm} 
f(u,v) \equiv \left(\begin{array}{c} \frac{1}{3} u^3 \cr v^3 \end{array} \right),
\eeq
for $u\in (-\infty, 3)$ and $v\in \R$. Their derivatives are
\beq \nn
dg(u,v) \equiv \left(\begin{array}{cc} u & 0 \cr 0 & v^2 \end{array} \right)  
\hspace{1cm} \text{and} \hspace{1cm} 
df(u,v) \equiv \left(\begin{array}{cc} u^2 & 0 \cr 0 & 3v^2 \end{array} \right),
\eeq
and the eigenvalue problem \eqref{eigenvalue-problem_right_mfd} has the solutions $r_1 = (1,0)$ with $\lambda_1=u$ and $r_2=(0,1)$ with $\lambda_2= 3$. Thus, this system is strictly hyperbolic on its domain   $(u,v) \in (-\infty, 3)\times \R$ and $dg$ has a non-trivial null space on the lines $\{u=0\}$ and $\{v=0\}$. Moreover, the first field is genuinely non-linear and the second linearly degenerate.
\end{Example}

\begin{Example} \label{example2}
\normalfont We modify the $p$-system in \cite{Smoller}, by considering 
\beq \label{example2_eqn0}
g(u,v) \equiv \left(\begin{array}{c} uv +v \cr \frac{1}{2}u^2v \end{array} \right)  
\hspace{1cm} \text{and} \hspace{1cm} 
f(u,v) \equiv \left(\begin{array}{c} u \cr u^2p(v) \end{array} \right),
\eeq
as accumulation and flux function in \eqref{system_cons_laws_manifold}. We assume $p(v)>0$ is a given smooth function and we restrict to $v>0$. In matrix form, \eqref{example2_eqn0} is given by
\beq \label{example2_eqn1}
\left( \begin{array}{cc} 1+u & v \cr \frac{1}{2}u^2 & uv \end{array} \right) \left( \begin{array}{c}  v_t \cr u_t \end{array} \right) \: + \: \left( \begin{array}{cc} 0 & 1 \cr u^2 p^\prime & 2up \end{array} \right) \left( \begin{array}{c}  v_x \cr u_x \end{array} \right) =0,  
\eeq
where $p^\prime$ is the derivative of $p=p(v)$. Clearly, $dg(v,u)$ has a non-trivial null-space at $u=0$ for all $v>0$. The resulting eigenvalue problem \eqref{eigenvalue-problem_right_mfd} is well-posed with eigenvalues
\beq \label{example2_eigenvalues}
\lambda_\pm = \frac{1}{2v(2+u)} \left( \alpha(v,u) \pm  \sqrt{8uv(2+u)p' + \alpha(v,u)^2}\right),
\eeq
for $\alpha(v,u)  \equiv 4p(v)(1+u)-u(1+2vp')$. Since at $u=0$
\beq \nonumber
\lambda_+ = \frac{2p(v)}{v} >0 \hspace{1cm} \text{and} \hspace{1cm} \lambda_- = 0,
\eeq
it follows that $\lambda_+$ and $\lambda_-$ are distinct for all $u$ sufficiently close to zero. Moreover, for $u$ sufficiently small, $\lambda_\pm$ are smooth functions for all $v>0$. Thus, since the flux and accumulation functions in \eqref{example2_eqn0} are smooth, the eigenvectors of $\lambda_\pm$ are smooth. To conclude that \eqref{example2_eqn1} is strictly hyperbolic in the sense of Definition \ref{def_hyperbolicity}, we need to choose $p(v)$ such that the resulting eigenvectors are linearly independent.      

For this, we here consider the simple (but non-trivial) special case $$p(v)=v.$$  The eigenvalues simplify to
$\lambda_+ = 2$ and $\lambda_- = - \frac{u}{(2+u)v}$ and the corresponding normalized eigenvectors are given by
\begin{eqnarray} \nonumber
r_+ &=& \frac1{\sqrt{(1-2v)^2+4(1+u)^2}}\left( \begin{array}{c}  1-2v \cr 2(1+u) \end{array} \right)  \cr
r_- &=& \frac1{\sqrt{4v^2+u^2}} \left( \begin{array}{c}  -2v \cr u \end{array} \right).
\end{eqnarray} 
Clearly $r_-$ and $r_+$ remain linearly independent in a neighbourhood of $u=0$, for suitable values of $v>0$. A computation shows that $(\lambda_-,r_-)$ is genuinely non-linear for all $u\neq0$ and all $v>0$, while $(\lambda_+,r_+)$ is linearly degenerate.
\end{Example}

\subsection{Weak Solutions and Rankine Hugoniot Conditions}

We now define weak solutions and derive the Rankine Hugoniot jump conditions for system \eqref{system_cons_laws_manifold}. As in \cite{Smoller}, we define $C^1_0(\Omega)$ to be the set of real valued continuously differentiable functions on $\Omega\equiv \{(x,t): t\geq 0\}$ with compact support (which can be non-vanishing on compact subsets of the line $\{t=0\}$). Now, multiplying \eqref{system_cons_laws_manifold} by some $\psi \in C^1_0(\Omega)$, integrating over the resulting equation and applying Gauss's Divergence Theorem to shift derivatives, leads to our definition of weak solutions.

\begin{Def} \label{weak_solution} 
Let $w(x,t)\in \M$ be a bounded function such that $f(w(x,t))$ and $g(w(x,t))$ are measurable in $(x,t)$ and let $w_0(x)\in \M$ be a bounded function such that $g(w_0(x))$ is measurable in $t$.
We say that $w(x,t)\in \M$ is a weak solution of \eqref{system_cons_laws_manifold} with initial data $w_0$ if 
\beq \nonumber
\iint\limits_{t\geq0} \big(g(w)\,\psi_t \,+\, f(w)\,\psi_x \big)\,dx dt\, + \int_\R \psi(\cdot,0)\, g(w_0)\,  dx \ =\ 0, \ \ \ \ \forall  \, \psi \in C^1_0(\Omega).
\eeq
\end{Def}

To clarify, the boundedness of $w(x,t)$ assumed in Definition \ref{weak_solution} shall be understood with respect to the canonical distance function $d_\M$, defined in \eqref{distancefunction}, which turns $\M$ into a metric-space. 

We next show that the discontinuity of a weak solution across a surface $\Gamma(t)\equiv (x(t),t) \in \Omega$ satisfies the following generalized \emph{Rankine Hugoniot jump conditions}:   
\beq \label{RHconditions}
s [g(w)] = [f(w)],
\eeq 
where $s(t)=\dot{x}(t)$ is the shock speed, and $[\cdot]$ denotes the jump across $\Gamma$, i.e., $[w]=w_l - w_r$, $[g(w)]=g(w_l)-g(w_r)$ and $[f(w)]=f(w_l)-f(w_r)$, with $w_l$ and $w_r$ denoting the left- and right-limit of $w(x,t)$ at $\Gamma$.

\begin{Prop} \label{weak-sol<=>RH-condition}
Let $\Gamma(t) \equiv (x(t),t)$ be a differentiable curve across which $w(x,t)$ is discontinuous and such that $w(x,t)$ is in $C^1$ on the complement of $\Gamma$ and $w(x,t)$ has well-defined limits on both sides of $\Gamma$. Suppose $w(x,t)$ is a strong solution of \eqref{system_cons_laws_manifold} in the complement of $\Gamma$. Then, $w(x,t)$ is a weak solution of \eqref{system_cons_laws_manifold} if and only if \eqref{RHconditions} hold across $\Gamma$.
\end{Prop}

\Proof    
The proof closely follows the reasoning in \cite{Smoller} and is only included for completeness. Consider some point $p$ on $\Gamma$ and let $B \subset \{(x,t): t\geq 0\}$ be some small open ball centered at $p$. Denote with $B_1$ and $B_2$ the two open regions in $B$ to the right and left of $\Gamma$, respectively. Now, assuming $w(x,t)$ is a weak solution and that $\Psi$ has compact support in $B$, we find from \eqref{weak_solution} that
\begin{eqnarray} \label{weak-sol<=>RH-condition_eqn1}
0 &=& \iint\limits_{B} \big(g(w)\psi_t + f(w)\psi_x \big) dx dt \cr
&=&   \iint\limits_{B_1} \big(g(w)\psi_t + f(w)\psi_x \big)  dx dt+  \iint\limits_{B_2} \big(g(w)\psi_t + f(w)\psi_x \big) dx dt. \ \ \ \ \
\end{eqnarray}
Using the fact that $w(x,t)$ is a strong solution of \eqref{system_cons_laws_manifold} on each $B_i$, $i=1,2$, since $w \in C^1(B_i)$, we get
\begin{eqnarray} \nonumber
\iint\limits_{B_i} \big(g(w)\psi_t + f(w)\psi_x \big) dx dt
&=& \iint\limits_{B_i} \big(g(w)\psi\big)_t + \big(f(w)\psi \big)_x dx dt.
\end{eqnarray}
Now, using the divergence theorem together with the fact that the outward pointing unit normal vector of $\partial B_1 =\Gamma$ and of $\partial B_2 =\Gamma$, $N_1$ and $N_2$, both point along the direction of $(-s,1)$, but with opposite sign, we write \eqref{weak-sol<=>RH-condition_eqn1} as      
\begin{eqnarray} \nonumber
\iint\limits_{B_1} \big(g(w)\psi_t + f(w)\psi_x \big) dx dt
&=& \int\limits_{\Gamma} \psi \left(\hspace{-0.2cm} \begin{array}{c} g(w_r) \cr f(w_r) \end{array} \hspace{-0.2cm}\right) \cdot N_1 \, dt , \cr
\iint\limits_{B_2} \big(g(w)\psi_t + f(w)\psi_x \big) dx dt
&=& -\int\limits_{\Gamma} \psi \left(\hspace{-0.2cm} \begin{array}{c} g(w_l) \cr f(w_l) \end{array} \hspace{-0.2cm}\right) \cdot N_1 \, dt .
\end{eqnarray}
Inserting this into \eqref{weak-sol<=>RH-condition_eqn1} and rescaling $N_1$ to coincide with $(-s,1)$, gives
\begin{eqnarray} \nonumber
0 &=& \iint\limits_{B} \big(g(w)\psi_t + f(w)\psi_x \big) = \int \psi \left(  s [g(w)] - [f(w)]  \right) dt ,
\end{eqnarray}
for all $\psi \in C^1_0$, from which we conclude that \eqref{RHconditions} holds if and only if $w(x,t)$ is a weak solution of \eqref{system_cons_laws_manifold}. 
\QED

\subsection{Left Eigen-One-Forms}

To prove Glimm's Theorem within our framework and to show the admissibility of shock waves, we use the following eigenvalue problem for eigen-one-forms $l_k$,         
\beq \label{eigenvalue-problem_left_mfd}
\lambda_k\ l_k (dg)  = l_k  (df) ,
\eeq
by which we mean that $l_k$ is a smooth mapping from $\M$ to the space of linear maps from $\R^n$ to $\R$, $\mathcal{L}(\R^n,\R)$. That is, $l_k|_w : \R^n \rightarrow \R$ is linear and smooth in $w\in \M$, and we have
\beq \nn
\lambda_k(w) \ l_k\ew \left(dg\ew(v)\right)  = l_k\ew \left(df\ew(v)\right), \ \ \ \ \ \ \forall \ v\in T_w\M.
\eeq
The above equation closes, since  $dg|_w(v) \in T_{g(w)}\R^n = \R^n$ and $df|_w(v) \in T_{f(w)}\R^n = \R^n$ for all $v \in T_w\M$, where we use the canonical identification between $T_p\R^n$ and $\R^n$, for all $p\in\R^n$, through the standard basis on $\R^n$. By a slight abuse of language we refer to the $l_k$'s as one-forms.  

Let us remark here that one can write the eigenvalue problem for eigen-one-forms \eqref{eigenvalue-problem_left_mfd} equivalently as the left-eigen\-vector problem used in \cite{Glimm,Smoller}. Namely, by the Riesz representation Theorem, there exists a unique vector $\vec{l}_k \in \R^n$ such that $l_k(v) = \langle \vec{l}_k ,v \rangle_{\R^n}$ for all $v \in \R^n$, where $\langle \cdot,\cdot \rangle_{\R^n}$ denotes the Euclidean inner product on $\R^n$.                   

The following lemma shows that the existence of left eigen-one-forms follows from the existence of right eigenvectors, under the assumption that $dg$ is point-wise invertible in $\M$.       We apply Lemma \ref{equivalence_left-right_lemma} in Section \ref{sec_Thm_mfd} to prove $C^2$ contact of the shock and rarefaction curves at $w_l$ and to prove the admissibility of the lower branch of shock curves.

\begin{Lemma} \label{equivalence_left-right_lemma}
Assume $\text{det}(dg|_w) \neq 0$ and that there exists $n$ linearly independent right eigenvectors $r_k(w) \in T_w\M$ which solve \eqref{eigenvalue-problem_right_mfd} with eigenvalues $\lambda_k(w) \in\R$, then there exists $n$ linearly independent one-forms, $l_k$, which solve \eqref{eigenvalue-problem_left_mfd}. In particular, there exist a normalization of the $l_k$'s such that the $l_k$'s and $r_k$'s are mutually orthonormal in the sense that
\beq \label{new_condition}
l_j\big(dg(r_k) \big) = \begin{cases} 1, \ \ j=k \cr 0, \ \ j\neq k   \end{cases}.
\eeq 
\end{Lemma}

\Proof                     
It suffices to prove the lemma at a fixed point $w\in\M$. 
Using that $dg|_w$ has the inverse   
\beq \nn
dg|_{g(w)}^{-1}  \ : \  \R^n \longrightarrow T_w\M,
\eeq 
where we identify $T_{g(w)}g(\M)$ with $\R^n$, we define the linear map 
\beq \nn
A(w) =  dg|_{g(w)}^{-1} df|_w ,
\eeq 
which maps $T_w\M$ into itself. Now, subsequently suppressing the dependence on $w$, we write \eqref{eigenvalue-problem_right_mfd} as 
\beq \label{equivalence_left-right_eqn1}
A \, r_k =  \lambda_k \, r_k .
\eeq  
Let $e_k$ be the $k$-th standard basis vector of $\R^n$, that is, all entries of $e_k$ are $0$ except the $k$-th entry, which equals $1$. Define the linear mapping $U\, : \R^n \longrightarrow T_w\M$ by setting $U(e_k) \equiv r_k$. Then $U$ is invertible and, in light of \eqref{equivalence_left-right_eqn1}, we obtain 
\beq \nn 
U^{-1}A U= \text{diag}(\lambda_1,...,\lambda_n),
\eeq
where we represent the linear mapping on the left hand by its matrix in the standard basis on $\R^n$. Multiplying the above equation from the left by $e_k^*$, where star denotes the transpose of a matrix, yields,
\beq \nn 
e_k^*\cdot U^{-1}A U= \lambda_k \, e_k^*,
\eeq
 which is equivalent to
\beq \nn
 e_k^*\cdot U^{-1} A = \lambda_k e_k^*\cdot U^{-1}.
\eeq
Substituting now the definition of $A$, that is, $A=dg^{-1} df$, into the above equation we obtain
\beq \nn
 e_k^*\cdot U^{-1} dg^{-1} df = \lambda_k\cdot e_k^*\cdot U^{-1} dg^{-1} dg.
\eeq
The left- and right-hand side of the above equation are both linear mappings from $T_w\M$ to $\R$, and applying them to some $v \in T_w\M$, yields
\beq \nn
 \Big\langle  \big(U^{-1}dg^{-1}\big)^* e_k,  df(v) \Big\rangle_{\R^n}
 = \lambda_k \cdot \Big\langle \big(U^{-1}dg^{-1}\big)^*e_k, dg(v) \Big\rangle_{\R^n} \ .
\eeq
We conclude that the sought after one-form, satisfying \eqref{eigenvalue-problem_left_mfd}, is given by
\beq \label{equivalence_left-right_eqn2}
l_k\ew(\cdot) = \Big\langle \big(U^{-1}dg^{-1}\big)^* e_k\, ,\, \cdot\, \Big\rangle_{\R^n} .
\eeq

To verify \eqref{new_condition}, we use \eqref{equivalence_left-right_eqn2} and compute
\begin{eqnarray} \nn
l_k\left(dg\big(r_j\big) \right) 
&=& \Big\langle \big(U^{-1}dg^{-1}\big)^* e_k\, ,\, dg(r_j) \Big\rangle_{\R^n}  \cr
&=& \Big\langle  e_k ,U^{-1}(r_j) \Big\rangle_{\R^n}  \cr
&=& \big\langle  e_k ,e_j \big\rangle_{\R^n},
\end{eqnarray}
which completes the proof.
\QED

A useful application of left eigen-one-forms, though not further used in this article, is the following Lemma (which is also proven in \cite{LambertMarchesin} for $\R^n$ as state space). 

\begin{Lemma} \label{what is genuine nonlinearity_lemma}
Assume that \eqref{system_cons_laws_manifold} is hyperbolic, that the flux $f$ and the accumulation $g$ are both $C^2$ regular and that $dg|_w$ is invertible for all $w\in\M$, then     
\beq \label{what is genuine nonlinearity}
 r_k\big(\lambda_k\big) \cdot l_k\Big( dg(r_k)\Big)  
= l_k\Big(\big(\nabla_{r_k} df\big)(r_k)\Big) - \lambda_k \cdot l_k\Big(\big(\nabla_{r_k} dg\big)(r_k)\Big),
\eeq
where $\nabla_{r_k} df$ and $\nabla_{r_k} dg$ are the covariant derivatives of the one forms $df^i$ and $dg^i$ for $i=1,...,n$ fixed.
\end{Lemma}

\Proof
This proof generalizes the one in \cite{Smoller}. To begin, note that the expressions on the left and right hand side of the $k$-th characteristic eigenvalue problem, \eqref{eigenvalue-problem_right_mfd}, are both differentiable maps from $\M$ to $\R^n$. Fixing Cartesian coordinates $x^i$ on $\R^n$, we differentiate the $i$-th component of \eqref{eigenvalue-problem_right_mfd} in the direction of $r_k$, which yields
\beq \nonumber
r_k\ew \big(\lambda_k\  dg^i\ew (r_k) \big) = r_k\ew \big(df^i\ew (r_k) \big),
\eeq
where $g^i \equiv x^i\circ g$ and $f^i \equiv x^i\circ f$, for $i=1,...,n$ fixed.
By the product rule of partial differentiation, applied for instance in some coordinates on $\M$, we write the above equation as
\beq \label{what is genuine nonlinearity_lemma_eqn0}
r_k\ew (\lambda_k)\  dg^i\ew (r_k)  + \lambda_k(w)\, r_k\ew \Big( dg^i\ew (r_k) \Big) = r_k\ew \Big(df^i\ew (r_k) \Big).
\eeq
Analogous to the computation in \eqref{shock-param-lemma_tech-eqn0c}, we find by the Leipnitz rule of covariant differentiation \eqref{covariant_derivtive_Leipnitz-rule} that
\beq \nn
r_k\ew \Big(df^i\ew (r_k) \Big) = \big( \nabla_{r_k} df^i\big)\ew (r_k) + df^i\ew\big(\nabla_{r_k}r_k\big),
\eeq
where $\nabla_{r_k} df^i$ denotes the covariant derivative of the one-form $df^i$ and where we used as in \eqref{shock-param-lemma_tech-eqn0c} that the Christoffel symbols enter covariant differentiation of vector fields and of one-forms with opposite sign, c.f. \eqref{covariant_derivative} and \eqref{covariant_derivative_oneforms}. Now, since an analogous result holds for $dg^i$, we write \eqref{what is genuine nonlinearity_lemma_eqn0} as
\begin{align} \nn
r_k\ew (\lambda_k)\  dg^i\ew (r_k)  + \lambda_k(w)\, \big( \nabla_{r_k} dg^i\big)\ew (r_k) + \lambda_k(w)\, dg^i\ew\big(\nabla_{r_k}r_k\big) \cr
= \big( \nabla_{r_k} df^i\big)\ew (r_k) + df^i\ew\big(\nabla_{r_k}r_k\big). \hspace{2cm}
\end{align}
Writing the $k$-th characteristic one-form in coordinates $x^i$ as $l_k\equiv (l_k)_i\, dx^i$, we multiply the previous equation by $(l_k)_i$ and sum over the resulting expression with respect to $i$, which yields
\begin{align} \label{what is genuine nonlinearity_lemma_eqn1}
r_k\ew (\lambda_k)\  l_k\Big(dg\ew (r_k) \Big) + \lambda_k(w)\, l_k\Big(\big( \nabla_{r_k} dg\big)\ew (r_k)\Big) + \lambda_k(w)\, l_k\Big(dg\ew\big(\nabla_{r_k}r_k\big)\Big) \cr
= l_k\Big(\big( \nabla_{r_k} df\big)\ew (r_k)\Big) + l_k\Big(df\ew\big(\nabla_{r_k}r_k\big)\Big). \hspace{2cm}
\end{align}
Now, from the eigen-value problem \eqref{eigenvalue-problem_left_mfd}, we find
\beq \nn
 \lambda_k(w)\, l_k\Big(dg\ew\big(\nabla_{r_k}r_k\big)\Big)  = l_k\Big(df\ew\big(\nabla_{r_k}r_k\big)\Big),
\eeq
so that \eqref{what is genuine nonlinearity_lemma_eqn1} simplifies to 
\beq \nn
r_k\ew (\lambda_k)\  l_k\Big(dg\ew (r_k) \Big) + \lambda_k(w)\, l_k\Big(\big( \nabla_{r_k} dg\big)\ew (r_k)\Big) 
= l_k\Big(\big( \nabla_{r_k} df\big)\ew (r_k)\Big),
\eeq
which is the sought after equation \eqref{what is genuine nonlinearity}, independent of any choice of coordinates. This completes the proof of Lemma \ref{what is genuine nonlinearity_lemma}.
\QED

\section{The Riemann Problem in the Manifold of States} \label{sec_Thm_mfd}

In this section, we extend Lax's existence theory \cite{Lax} to the geometric framework introduced in Section \ref{sec_frame}, for which we use mostly the exposition in \cite{Smoller} as a starting point. As in \cite{Lax}, the solutions of Riemann problems constructed are compositions of contact discontinuities, shock and rarefaction waves.

\subsection{Rarefaction Waves} \label{sec_Thm_mfd_rarefaction-waves}

We begin by introducing the $k$-th characteristic curve of \eqref{system_cons_laws_manifold} as the (real valued) solution of the ODE
\beq \label{characteristic_def}
\frac{d}{dt} \chi(x,t) = \lambda_k\big(w(\chi(x,t),t)\big), \hspace{.5cm} \text{with} \hspace{.5cm} \chi(x,0)=x.
\eeq 
Following \cite{Dafermos}, we now define rarefaction waves as $C^1$ solutions of \eqref{system_cons_laws_manifold} constant along a family of characteristic curve and such that the corresponding wave speed increase monotonously with respect to $x$, so that the family of characteristic curves diverge to form a fan-like region.

\begin{Def} \label{simple_wave_def}
We say that a $C^1$ solution of \eqref{system_cons_laws_manifold}, $(x,t) \mapsto w(x,t) \in \M$, is a $k$-simple wave, if it is constant along the $k$-th characteristic curve, that is, if for fixed $x$ the tangent vector of the curve $t \mapsto w\big(\chi(x,t),t\big)$ satisfies
\beq\nn
\frac{d}{dt} w\big(\chi(x,t),t\big) = 0 .
\eeq
A $k$-rarefaction wave is a $k$-simple wave with $\frac{\partial}{\partial x} \lambda_k\big(w(x,t)\big) >0$ and a $k$-compression wave is a $k$-simple wave with $\frac{\partial}{\partial x} \lambda_k\big(w(x,t)\big) <0$.
\end{Def}

It follows immediately that characteristic curves are straight lines for simple waves. 

Under the assumption that the $k$-th characteristic field is genuinely nonlinear, the following theorem proves that for each $w_l \in \M$ there exists a one-parameter family of states $w_r$ for which the Riemann problem with $w_l$ on the left and $w_r$ on the right is solved by a rarefaction wave.

\begin{Thm} \label{Thm_simple-waves}
Assume \eqref{system_cons_laws_manifold} is strictly hyperbolic, let $w_l$ be some point in $\M$ and assume the $k$-th characteristic field of \eqref{system_cons_laws_manifold} is genuinely nonlinear in some neighborhood of $w_l$ with $r_k\ew\big(\lambda_k \big)>0$. Then there exists a one-parameter family of states $w(\epsilon)$, with $\epsilon \in [0,a)$ for some $a>0$, such that $w(0)=w_l$ and $w(\epsilon)$ can be connected to $w_l$ on the right by a $k$-rarefaction wave in $\M$ for each $\epsilon \in [0,a)$. We call the curve $\epsilon \mapsto w(\epsilon)$ the $k$-th rarefaction curve. This curve is unique modulo its parameterization and it can be parameterized so that its tangent vector is given by $\dot{w}(\epsilon)=r_k(w(\epsilon))$. 
\end{Thm}

\Proof
Consider the ODE
\begin{eqnarray} \label{Thm_simple-waves_ODE}
\frac{d v}{d\epsilon} = {r}_k\big(v(\epsilon) \big)  \hspace{.5cm} \text{with} \hspace{.5cm} 
v(\lambda_k(w_l)) = w_l.
\end{eqnarray}
Since $df$ and $dg$ are assumed $C^1$ regular, we conclude that $r_k$ and $\lambda_k$ are $C^1$ as well. Now the Picard Lindel\"of theorem yields the existence of a constant $a>0$ and a function $v(\epsilon)$, for $\epsilon \in [\lambda_k(w_l),\lambda_k(w_l)+a)$, which solves \eqref{Thm_simple-waves_ODE}. 
Consider now the following Riemann problem for the scalar conservation law
\beq \label{Thm_simple-waves_eqn1}
\epsilon_t + \lambda_k\big( v(\epsilon)\big)\, \epsilon_x =0, \hspace{1cm}  \epsilon(x,0)= \begin{cases} \lambda_k(w_l), \ \ \ \ \ \ \ x<0 \cr \lambda_k(w_l) + \hat{a}, \ \ x>0,   \end{cases}
\eeq
for some $0< \hat{a} < a$. By genuine non-linearity of the $k$-th characteristic field, we find that
\begin{eqnarray} \nonumber
\frac{d}{d\epsilon} \lambda_k\big(v(\epsilon)\big) 
=  d\lambda_k\bigg\rvert_{v(\epsilon)} \hspace{-.2cm} \left(\frac{dv}{d\epsilon}\right) = {r}_k\bigg\rvert_{v(\epsilon)} \hspace{-.4cm}\big(\lambda_k \big) >0 ,
\end{eqnarray}
which implies $\lambda_k\circ v$ to be monotone in $\epsilon$. Thus, following the basic theory of \emph{scalar} conservation laws \cite{Smoller}, there exist a rarefaction wave solving \eqref{Thm_simple-waves_eqn1}, given by
\beq \label{Thm_simple-waves_eqn2}
\epsilon(x,t) = (\lambda_k\circ v)^{-1}\left(\frac{x}{t}\right),
\eeq 
for $(x,t)$ lying in the wave fan region 
\beq \label{wavefan}
\lambda_k(w_l) \ \leq \ \frac{x}{t} \ \leq \ \lambda_k\Big(v\big(\lambda_k(w_l) + \hat{a}\big)\Big),
\eeq
where we used the initial data in \eqref{Thm_simple-waves_ODE} to simplify the lower bound in \eqref{wavefan}. Notice that $\epsilon(x,t)$ is constant along the characteristic lines of \eqref{Thm_simple-waves_eqn1}, (the straight lines inside the wave fan \eqref{wavefan}), and its range is the domain of definition of $v$, that is, $[\lambda_k(w_l),\lambda_k(w_l)+\hat{a}]$.  This allows us to define  $(x,t)\mapsto w(x,t)$ as  
\beq \label{Thm_simple-waves_def_of_soln}
w(x,t) \equiv v\big(\epsilon(x,t)\big),
\eeq
for $(x,t)$ lying in the wave fan \eqref{wavefan}. Since $\epsilon(x,t)$ is constant along the $k$-characteristic curves and $C^1$ regular inside the wave fan \eqref{wavefan} and Lipschitz continuous across its boundary lines, we conclude that $w$ is a $k$-simple wave provided it solves \eqref{system_cons_laws_manifold}-\eqref{Riemann_problem}. 

We now show that $w$ is a solution of the Riemann problem \eqref{Riemann_problem} for the system \eqref{system_cons_laws_manifold}. For this, we compute
\begin{eqnarray} \nonumber
g(w)_t + f(w)_x 
&=& dg\ew (w_t) +df\ew (w_x) \cr
&=& dg \ew \left(\frac{dv}{d\epsilon} \right)\epsilon_t + df \ew \left(\frac{dv}{d\epsilon}\right) \epsilon_x ,
\end{eqnarray}
and using \eqref{Thm_simple-waves_ODE} together with the eigenvalue problem \eqref{eigenvalue-problem_right_mfd}, we find
\begin{eqnarray} \label{Thm_simple-waves_eqn0}
g(w)_t + f(w)_x 
&=& dg \ew \left({r}_k(v) \right)\epsilon_t + df \ew \left({r}_k(v)\right) \epsilon_x \cr
&=& dg \ew \left({r}_k(v) \right)  \left(\epsilon_t + \lambda_k(v) \epsilon_x \right).
\end{eqnarray}
From \eqref{Thm_simple-waves_eqn1}, it follows that the right hand side in the previous equation vanishes, from which we conclude that $w$ is in fact a solution of \eqref{system_cons_laws_manifold} and that $w$ is a simple wave. 

Finally, use $\lambda_k(w)=\lambda_k(v(\epsilon))$ and \eqref{Thm_simple-waves_eqn2} to compute
\beq\nn
\frac{\partial}{\partial x} \lambda_k(w) = \frac{1}{t} >0,
\eeq
which implies that $w(x,t)$ is indeed a rarefaction wave. 

Varying now $\hat{a}\in (0,a)$ yields the sought after one-parameter family of states connected to $w_l$ by a rarefaction wave.  Uniqueness of the rarefaction curve follows from the uniqueness of $r_k$ (modulo its length) under the assumption that \eqref{system_cons_laws_manifold} is strictly hyperbolic.
\QED

Let us remark that if  the set of points $\Sigma$, for which the null-space of $dg$ is non-trivial, were not a set of measure zero, then the eigenvalues $\lambda_k$ would not be specified over this set and \eqref{Thm_simple-waves_eqn1} could not be introduced. Let us remark further, that one could avoid the inversion problem $(\lambda_k\circ v)^{-1}$ in \eqref{Thm_simple-waves_eqn2} by normalizing $r_k$ such that ${r}_k\ew\big(\lambda_k \big)=1$ holds, however, we avoid this normalization here since it would prevent us from introducing an arc-length parameterization for the wave curves, which is convenient for extending Glimm's Theorem to our framework.

\subsection{Shock Waves} \label{sec_Thm_mfd_shock-waves}

We now prove the existence of $n$ one-parameter families of states that can be connected to $w_l$ on the right by a shock wave. 

\begin{Thm} \label{shock curves_Thm}
Let $w_l \in \M$, and assume that \eqref{system_cons_laws_manifold} is strictly hyperbolic in some neighborhood of $w_l$. Then there exists $n$ one-parameter families of states, $\epsilon \mapsto w_k(\epsilon) \in C^2(I,\M)$ with $I\equiv (-a,a)$ for some $a>0$, such that $w_k(0)=w_l$ and $w_k(\epsilon)$ satisfies the RH jump conditions \eqref{RHconditions} for some unique shock speed $s_k \in C^1(I,\R)$, for all $\epsilon \in I$. The shock curves $\epsilon \mapsto w_k(\epsilon)$ are linearly independent at $w_l$, unique up to parameterization and have a $C^2$-dependence on $w_l$. 
\end{Thm}    

\Proof
Given $w_l\in \M$, we choose coordinates $y^\mu$ defined in a neighborhood $\mathcal{U}$ of $w_l$, such that any straight line in $\R^n$ through $y(w_l)$ intersects $y(\Sigma\cap \mathcal{U})$ at most in finitely many points.\footnote{That is, we arrange for each hypersurface in $y(\Sigma\cap \mathcal{U})$ to be sufficiently curved.} 
Writing $y= y(w) \in \R^n$ and $y_l= y(w_l) \in \R^n$, we compute
\begin{eqnarray}\nonumber
f(w) - f(w_l) 
&=& f\circ y^{-1} (y) -  f\circ y^{-1} (y_l)    \cr
&=& \int\limits_0^1 \frac{d}{d\sigma} f\circ y^{-1}\big( y_l + \sigma (y - y_l) \big) d\sigma \cr
&=& \int\limits_0^1 d(f\circ y^{-1})\big|_{ y_l + \sigma (y - y_l)} d\sigma \cdot (y - y_l) \cr
&\equiv & \mathcal{F}(y) \cdot (y - y_l)
\end{eqnarray}
and similarly
\begin{eqnarray}\nonumber
g(w) - g(w_l) 
&=& \int\limits_0^1 d(g\circ y^{-1})\big|_{ y_l + \sigma (y - y_l)} d\sigma \cdot (y - y_l) \cr
&\equiv & \mathcal{G}(y) \cdot (y - y_l) .
\end{eqnarray}
We now write the RH condition \eqref{RHconditions} as the equivalent eigenvalue problem
\beq \label{RHconditions_integrated}
\big( \mathcal{F}(y) - s\, \mathcal{G}(y) \big) (y-y_l) = 0.
\eeq

Assume for the moment that $w_l \notin \Sigma$, i.e.,  $dg|_{w_l}$ is invertible. 
At $y_l$, the above matrices simplify to $\mathcal{F}(y_l) = d\left(f\circ y^{-1}\right)|_{y_l}$ and $\mathcal{G}(y_l) = d\left(g\circ y^{-1}\right)|_{y_l}$ so that \eqref{RHconditions_integrated} reduces to the right eigenvalue problem \eqref{eigenvalue-problem_right_mfd} at $y_l$. Thus, for $y$ sufficiently close to $y_l$, the characteristic polynomial $\mathcal{P}$ of \eqref{RHconditions_integrated} is a small perturbation of the characteristic polynomial of \eqref{eigenvalue-problem_right_mfd} at $y_l$. By continuous dependence of the (complex) zeros of a polynomial on its coefficients, we conclude that $\mathcal{P}$ has $n$ mutually distinct zeros $\mu_i$, $i=1,...,n$, with their real parts satisfying 
\beq \label{shock curves_Thm_techeqn0}
\text{Re}(\mu_1) > ... > \text{Re}(\mu_n).
\eeq 
Now, since $\mathcal{P}$ has real coefficients, the genuine complex roots of $\mathcal{P}$ appear in complex conjugate pairs, so that \eqref{shock curves_Thm_techeqn0} requires $\mathcal{P}$ to have $n$ real valued zeros. Therefore, \eqref{RHconditions_integrated} has $n$ real valued eigenvalues $\mu_1 > ... > \mu_n$, that is, there exists $\zeta_k(y) \in \R^n$ and $\mu_k(y) \in \R$ such that
\beq \label{shock curves_Thm_techeqn1} 
\big( \mathcal{F}(y) -  \mu_k(y) \mathcal{G}(y) \big) \zeta_k(y) = 0,
\eeq
for $k=1,...,n$, and where  
\beq \nn
\zeta_k(y_l)=dy\Big\rvert_{w_l}\hspace{-.3cm}\big(r_k\big) \ \ \ \ \ \ \ \text{and} \ \ \ \ \ \ \ 
\mu_k(y_l)= \lambda_k(y_l). 
\eeq

Now, assuming $dg|_{w}$ has a non-trivial null-space for all $w\in \Sigma$ and $\Sigma \neq \emptyset$, then the characteristic polynomial of \eqref{eigenvalue-problem_right_mfd} vanishes identically at each $w\in \Sigma$. However, since by our choice of coordinates $y$, we arranged for $y(\Sigma\cap \mathcal{U})$ to be intersected by the straight line $\sigma \mapsto y_l + \sigma (y - y_l)$ at most in finitely many points, it follows that $\mathcal{G}$ is invertible, so that the characteristic polynomial of \eqref{RHconditions_integrated} is non-vanishing and again gives rise to $n$ distinct real eigenvalues, that is, to \eqref{shock curves_Thm_techeqn1}. 

We conclude that \eqref{RHconditions_integrated} is satisfied if and only if
\begin{eqnarray} \label{RHconditions_integrated_aux}
y-y_l = \epsilon\, \zeta_k(y) \ \ \ \ \ \text{and} \ \ \ s= \mu_k(y),
\end{eqnarray}
for some non-zero $\epsilon \in \R$ and some $k=1,...,n$. It remains to prove that for each $k$ there exists a one-parameter family of states $y(\epsilon)$ which solves 
\beq \nn
H(y,\epsilon)\equiv y-y_l - \epsilon\, \zeta_k(y) =0.
\eeq
$H$ is a $C^2$ function since $g$ and $f$ are assumed to be in $C^3$. It is straightforward that $H(y_l,0) = 0$ and
\beq \nn
\frac{\partial}{\partial y^\nu}H^\mu\Big|_{(y_l,0)} = I^\mu_\nu ,
\eeq 
for $I$ denoting the identity matrix on $\R^n$ and $\mu,\nu \in \{1,...,n\}$ indices. Thus, the Implicit Function Theorem implies the existence of a $C^2$ curve $y_k(\epsilon)$, such that 
$$
H\big(y_k(\epsilon),\epsilon\big)=0
$$ 
for all $\epsilon$ sufficiently close to $0$.     

Defining $w_k(\epsilon) \equiv y^{-1}\big(y_k(\epsilon)\big)$, it is immediate by construction that $w_k$ is the sought after shock curve of the $k$-th characteristic field through $w_l$ with $s_k(\epsilon) \equiv \mu_k\circ y\,\big(w_k(\epsilon)\big)$ being the corresponding shock speed. Since $\M$ is assumed to be a $C^3$ manifold, we have $w_k \in C^2(I_k,\M)$. The above construction yields $n$ unique shock curves with the property that $w_k(0)=w_l$ and $\dot{w}(0)=r_k(w_l)$, so that linear independence follows as well.  The shock curves have a $C^2$-dependence on $y_l$, since $H$ has this dependence on $y_l$.

Finally, the above construction of shock curves and speeds is independent of the choice of coordinates, since $s(w)$, $[g^i(w)]$ and $[f^i(w)]$ are scalar functions over $\M$, c.f. Appendix \ref{sec_preliminaries}. 
\QED

The next lemma shows that the shock and rarefaction curves have $C^2$ contact at $w_l$, most remarkably, when parameterized by their \emph{arc-length}.  This lemma is the first step in the construction which requires Riemannian Geometry over basic Differential Topology.

\begin{Lemma} \label{shock-param-lemma}
Assume the hypotheses of Theorem \ref{shock curves_Thm}.  Assume further that $\M$ is endowed with a Riemannian metric $\langle\cdot,\cdot\rangle_\M$ and that the characteristic fields are normalized such that 
\beq \label{normalization_r_k}
\langle r_k,r_k\rangle_\M(w)=1, \ \ \ \ \ \forall\, w\in\M.
\eeq 
Then the shock curves defined in Theorem \ref{shock curves_Thm} can be parameterized by arc-length and, denoting the $k$-th shock curve in its arc-length parameterization by $\epsilon \mapsto  w_k(\epsilon)$, they satisfy 
\beq \label{shock-param-lemma_curves}
\dot w_k(0) = r_k(w_l) \ \ \ \ \ \ \  \text{and} \ \ \ \ \  \ddot w_k(0) = \nabla_{r_k}r_k(w_l),
\eeq
while the shock speeds satisfies 
\beq \label{shock-param-lemma_speeds}
s_k(0) = \lambda_k(w_l) \ \ \ \ \ \ \ \ \text{and} \ \  \ \ \ \ \dot{s}_k(0) = \frac12  r_k\Big|_{w_l}\hspace{-.3cm}(\lambda).
\eeq
Here ``$\cdot$'' denotes $\tfrac{d}{d\epsilon}$, $\nabla$ is the Levi Civita connection of $\langle\cdot,\cdot\rangle_\M$ and $\ddot w_k$ denotes covariant differentiation of $\dot{w}_k$ along the shock curve with respect to $\nabla$, c.f. \eqref{covariant_derivative_curve}.
\end{Lemma}

\Proof
It suffices to prove the theorem for $w_l \notin \Sigma$, since the $\lambda_k$'s and $r_k$'s are assumed to be extendable to $C^2$ functions over $\M$ (by Definition \ref{def_hyperbolicity}), so that the expressions on the right hand side of \eqref{shock-param-lemma_curves} and \eqref{shock-param-lemma_speeds} all extend  continuously to $\Sigma$ as $w_l$ varies, and since the shock curves have a $C^2$ dependence on $w_l$.

By construction of $w_k$ and $s_k$ we already have $w_k(0)=w_l$ and $s_k(0) = \lambda_k(w_l)$. Let $y^\mu$ be the coordinate system used in the proof of Theorem \ref{shock curves_Thm}. Writing $w_k(\epsilon') \equiv y^{-1}\big(y_k(\epsilon')\big)$, for some parameter $\epsilon'$, we find               
\beq \nn
\dot{w}_k(0) = dy^{-1}\bigg\rvert_{y_l} \big( \dot{y}_k(0)\big) \ \ \ \ \ \in \ T_{w_l} \M .
\eeq
Differentiating the first equation in \eqref{RHconditions_integrated_aux} with respect to $\epsilon'$ yields
\beq \nn
\dot y_k(0) = \zeta_k(y_l)    =  dy\ewl \big(r_k\big).
\eeq 
Combining the above two equations, we obtain
\beq \label{shock-param-lemma_tech-eqn0}
\dot{w}_k(0) = r_k(w_l)  ,
\eeq
which holds independent of the choice of coordinates, as both sides are objects in $T_{w_l} \M$.

Now, since $\dot{w}_k(0) = r_k(w_l)\neq 0$ we can use the change of variables defined in \eqref{arc-length_change}, that is, 
\beq\nn
\frac{d\epsilon}{d\epsilon'} = \sqrt{\left|\frac{d w_k}{d\epsilon'}\right|_\M} \hspace{.5cm} \text{with} \hspace{.5cm} \epsilon(0)=0,
\eeq 
to obtain the arc-length parameter $\epsilon$ of the shock curves. By \eqref{shock-param-lemma_tech-eqn0} and since $r_k$ is normalized by  $\langle r_k,r_k\rangle_\M=1$, we have $\tfrac{d\epsilon}{d\epsilon'}=|\dot{w}_k(0)|_\M=1$. Thus the shock curves parameterized by arc-length $\epsilon$ still satisfies $w_k(0)=w_l$ and \eqref{shock-param-lemma_tech-eqn0} with ``$\cdot$'' now denoting $\tfrac{d}{d\epsilon}$.

To derive the remaining identities, first differentiate the RH conditions \eqref{RHconditions} with respect to $\epsilon$, which gives 
\beq \label{shock-param-lemma_tech-eqn0b}
\dot{s} \big( g(w_k) - g(w_l) \big) + s dg \big(\dot{w}_k\big) = df \big(\dot{w}_k \big),
\eeq
where $dg$ and $df$ are evaluated at $w_k(\epsilon)$. We next differentiate \eqref{shock-param-lemma_tech-eqn0b} by $\epsilon$, for this observe that $df^i(\cdot)$ and $dg^i(\cdot)$ for fixed $i=1,...,n$ are one-forms on $\M$, so that, setting $a_\mu \equiv \frac{\partial f^i\circ x^{-1}}{\partial x^\mu}$ and $x^\mu(\epsilon) \equiv x^\mu\circ w_k(\epsilon)$ for some coordinate system $x^\mu$ and using the definition of $df$ in \eqref{differential}, we find
\begin{eqnarray} \label{shock-param-lemma_tech-eqn0c}
\frac{d}{d\epsilon} df^i\bigg|_{w_k}\hspace{-.2cm}\big(\dot{w}_k \big) 
&=& \frac{d}{d\epsilon} \left( \frac{\partial f^i\circ x^{-1}}{\partial x^\mu}  \frac{d x^\mu}{d\epsilon}  \right)  \cr
&=& \frac{\partial a_\mu}{\partial x^\nu} \,\frac{d x^\nu}{d\epsilon} \frac{d x^\mu}{d\epsilon}  
\,+\, a_\mu \,  \frac{d^2 x^\mu}{d\epsilon^2}  \cr
&=& \frac{\partial a_\mu}{\partial x^\nu} \,\dot{w}_k^\nu \dot{w}_k^\mu 
- \Gamma^\sigma_{\mu\nu} \, a_\sigma  \, \dot{w}_k^\nu \dot{w}_k^\mu 
\,+\, a_\mu \,  \frac{d \dot{w}_k^\mu}{d\epsilon} 
+ a_\sigma  \, \Gamma^\sigma_{\mu\nu} \, \dot{w}_k^\nu \dot{w}_k^\mu  \cr
&=&  \big(\nabla_{\frac{\partial}{\partial \epsilon}} a_\mu\big) \dot{w}_k^\mu  
\ + \ a_\mu \big( \nabla_{\frac{\partial}{\partial \epsilon}} \dot{w}_k^\mu \big) \cr 
&=&  \big(\nabla_{\frac{\partial}{\partial \epsilon}}df^i \big)\big(\dot{w}_k\big)  + df^i(\ddot{w}_k\big), 
\end{eqnarray}     
where $\ddot{w}_k$ denotes covariant differentiation along the shock curves, $\nabla_{\frac{\partial}{\partial\epsilon}}\dot w_k$. (Regarding the above cancellation, note that $\Gamma^\mu_{\sigma\rho}$ enters covariant differentiation of a vector field with a plus sign, c.f. \eqref{covariant_derivative}, but  covariant differentiation of a one-form with a minus sign, c.f. \eqref{covariant_derivative_oneforms}.) An analogous result holds for $dg^i$. Using now \eqref{shock-param-lemma_tech-eqn0c} for differentiating \eqref{shock-param-lemma_tech-eqn0b} by $\tfrac{d}{d\epsilon}$ component-wise then results in  
\begin{eqnarray} \nn
& & \ddot{s}  \big( g^i(w_k) - g^i(w_l) \big) + 2\dot{s} dg^i (\dot{w}_k) + s dg^i(\ddot{w}_k)+ s \nabla_{\frac{\partial}{\partial \epsilon}}dg^i \big(\dot{w}_k\big) \cr 
&=& \nabla_{\frac{\partial}{\partial \epsilon}}df^i \big(\dot{w}_k\big)  + df^i(\ddot{w}_k\big) ,
\end{eqnarray}
Evaluating this equation at $\epsilon =0$ and using that $w_k(0)=w_l$ and $\dot{w}_k(0) = r_k(w_l)$, yields              
\begin{eqnarray} \label{shock-param-lemma_tech-eqn1}
 & & 2\dot{s}(0) dg^i\ewl (r_k) + \lambda_k(w_l) dg^i\ewl(\ddot{w}_k(0)) + \lambda_k(w_l)  \big(\nabla_{r_k} dg^i\big)\ewl\big(r_k\big) \cr
 &=& \big(\nabla_{r_k}df^i\big)\ewl\big(r_k\big)  + df^i\ewl(\ddot{w}_k(0)\big) .
\end{eqnarray}
Differentiate now the eigenvalue problem \eqref{eigenvalue-problem_right_mfd} along $w_k$, which gives us
\beq \nn
\dot{\lambda}_k\,  dg^i \big(r_k\big) + \lambda_k  dg^i \big(\nabla_{\frac{\partial}{\partial \epsilon}}r_k\big) + \lambda_k \left(\nabla_{\frac{\partial}{\partial \epsilon}}dg^i\right) \big(r_k\big)     =      df^i \big(\nabla_{\frac{\partial}{\partial \epsilon}}r_k\big)  +   \left(\nabla_{\frac{\partial}{\partial \epsilon}} df^i \right)\big(r_k\big)  .
\eeq
Evaluating the previous equation at $\epsilon =0$, using $\dot{w}_k(0) = r_k(w_l)$, and subtracting the resulting expression from \eqref{shock-param-lemma_tech-eqn1}, yields
\beq \label{shock-param-lemma_tech-eqn2}
\big( 2\dot{s}(0) - \dot{\lambda}_k(0) \big) dg\ewl (r_k) 
 = \left( df -  \lambda_k\, dg \right)\hspace{-.1cm}\ewl  \big(\ddot{w}_k(0) - \big(\nabla_{r_k}r_k\big)(w_l)\big) ,
\eeq
where we used that $\dot{w}_k(0) = r_k(w_l)$ implies $ \big(\nabla_{\frac{\partial}{\partial \epsilon}}r_k\big)(w_k(0)) =\nabla_{r_k}r_k(w_l)$.

Now, in light of the eigenvalue problem \eqref{eigenvalue-problem_left_mfd} and since the left and right hand sides in \eqref{shock-param-lemma_tech-eqn2} are both elements in $T_{w_l} \M$, applying the eigen-one-form $l_k$ to both sides of \eqref{shock-param-lemma_tech-eqn2} yields
\beq \nn
\big( 2\dot{s}(0) - \dot{\lambda}_k(0) \big)\, l_k\Big(dg\ewl (r_k)\Big) = 0.
\eeq
Using that $l_k\Big(dg\ewl (r_k)\Big) =1$ by \eqref{new_condition} of Lemma \ref{equivalence_left-right_lemma} and that 
\begin{eqnarray} \label{shock-param-lemma_tech-eqn2b}
\dot{\lambda}_k(0) =  \frac{d}{d\epsilon}\bigg\rvert_{\epsilon=0}\hspace{-.4cm} \lambda_k\big(w_k(\epsilon)\big) 
= d\lambda_k\ewl \big(\dot{w}_k(0) \big) 
=  d\lambda_k\ewl \big(r_k\big) = 
r_k\ewl (\lambda_k)  ,
\end{eqnarray}
it follows that
\beq \label{shock-param-lemma_tech-eqn3}
\dot{s}(0) = \frac12  r_k\ewl(\lambda_k),
\eeq 
which proves the second equation in \eqref{shock-param-lemma_speeds}. 

To complete the proof, substitute \eqref{shock-param-lemma_tech-eqn3} into \eqref{shock-param-lemma_tech-eqn2}, which gives us
\beq \nn
\left( df -  \lambda_k\, dg \right)\hspace{-.1cm}\ewl  \big(\ddot{w}_k(0) - \nabla_{r_k}r_k(w_l)\big) =0,
\eeq
from which we conclude in light of the eigenvalue problem \eqref{eigenvalue-problem_right_mfd} that
\beq \label{shock-param-lemma_tech-eqn4}
\ddot{w}_k(0) - \nabla_{r_k}r_k(w_l) = c\, r_k(0) .
\eeq
Now, since $r_k$ is assumed to satisfy the normalization \eqref{normalization_r_k}, using the Leipnitz rule for $\nabla$, we find that
\beq \nn
0= r_k\ew\big(\langle r_k,r_k\rangle_\M\big) = 2 \langle \nabla_{r_k}r_k,r_k\rangle_\M(w).
\eeq
Likewise, since $\epsilon \mapsto w_k(\epsilon)$ is parameterized by arc-length, it follows by \eqref{arc-length_acceleration} that the acceleration $\ddot{w}_k$ is orthogonal to its velocity $\dot{w}_k$,  $\langle\ddot{w}_k,\dot{w}_k \rangle_\M=0$. Thus, applying the one-form $\langle\cdot,\dot{w}_k \rangle_\M$ to both sides of \eqref{shock-param-lemma_tech-eqn4} and using that $\dot{w}_k(0)=r_k(w_l)$, we conclude that $c=0$ and therefore \eqref{shock-param-lemma_tech-eqn4} implies the sought after equation \eqref{shock-param-lemma_curves}. 
\QED

As for a criteria guaranteeing uniqueness of solutions of the Riemann problem for the system \eqref{system_cons_laws_manifold}, we impose the so-called \emph{Lax admissibility} conditions. That is, we say that the $k$-th shock curve is (Lax) admissible if the following inequalities hold
\begin{eqnarray} \label{Lax_conditions}
\lambda_{k-1}(0) &<& s_k(\epsilon) < \lambda_k(0), \cr
 \lambda_{k}(\epsilon) &<& s_k(\epsilon) < \lambda_{k+1}(\epsilon) .
\end{eqnarray}
This condition is identical to the flat case and implies that the $k$-th characteristic lines impinge on the $k$-th shock, c.f. \cite{Lax,Smoller}. 

The next lemma shows that the shock curves in the arc-length parameterization of Lemma \ref{shock-param-lemma} are admissible for $\epsilon<0$. Its proof is taken from \cite{Smoller} and is presented here for completeness.

\begin{Lemma}
Assume the hypotheses of Lemma \eqref{shock-param-lemma} and let the $k$-th shock curve be parameterized by arc-length. Assume further that the $k$-th characteristic field is genuinely nonlinear with $r_k\ew(\lambda_k)>0$. Then \eqref{Lax_conditions} holds for $s_k(\epsilon)$ and $\lambda_k(w_k(\epsilon))$ if and only if $\epsilon <0$.
\end{Lemma}
\Proof
Define the function $\Phi(\epsilon)\equiv \lambda_k(\epsilon) - s_k(\epsilon)$, then Lemma \ref{shock-param-lemma} implies that $\Phi(0)=0$ and 
\beq\nn
\dot{\Phi}(0) = \dot{\lambda}_k(0) - \dot{s}_k(0) = \frac12  r_k\ewl(\lambda) >0 ,
\eeq
where we used \eqref{shock-param-lemma_tech-eqn2b} for the substitution $\dot{\lambda}_k(0)=r_k\ewl(\lambda_k)$.  Assume now that the Lax admissibility conditions hold. This implies that $\Phi(\epsilon)<0$ and thus that $\epsilon<0$, since $\Phi$ is monotonically increasing by the previous computation.

Conversely, assuming that $\epsilon <0$, the monotonicity of $\Phi$ implies that $\Phi(\epsilon)<0$ and thus $\lambda_k(\epsilon) <s_k(\epsilon)$. Moreover, since $\lambda_k(0)=s_k(0)$ and $\dot{s}_k(0)=\frac12 r_k\ewl(\lambda) >0$, we conclude that $\lambda_k(0) > s_k(\epsilon)$ for $\epsilon$ sufficiently close to $0$. Furthermore, as $\epsilon$ approaches $0$ from below, $s_k(\epsilon)$ converges to $\lambda_k(0)$, which implies that $s_k(\epsilon)>\lambda_{k-1}(0)$, since $\lambda_k(0)>\lambda_{k-1}(0)$ and $\lambda_k(0) > s_k(\epsilon)$. Finally, since $\lambda_{k+1}(0) > \lambda_k(0) = s_k(0)$, we conclude that $\lambda_{k+1}(\epsilon) > s_k(\epsilon)$ for $\epsilon <0$ sufficiently close to $0$. In summary, we proved that \eqref{Lax_conditions} holds.
\QED

Following \cite{Lax}, we now define the \emph{wave curve} for the genuinely non-linear characteristic fields as  rarefaction curves whenever $\epsilon>0$ and shock curves for $\epsilon<0$. 

\begin{Def}
Assume the $k$-th characteristic field is genuinely nonlinear with $r_k\ew(\lambda_k)>0$ and assume $r_k$ is normalized by \eqref{normalization_r_k}. Let $\bar{w}_k$ denote the $k$-th shock curve parameterized by arc-length and let $\hat{w}_k$ denote the $k$-th rarefaction curve constructed in Theorem \ref{Thm_simple-waves}. We now define the ``wave'' curve $\epsilon \mapsto w_k(\epsilon)$ in $\M$ as 
\beq \label{curve_state-space}
w_k(\epsilon) \equiv \begin{cases} \bar{w}_k(\epsilon), & \epsilon \leq 0 \\  \hat{w}_k(\epsilon), & \epsilon \geq 0  . \end{cases} 
\eeq
\end{Def}

Let us finally record the following basic properties of the wave curves being $C^2$ regular and parameterized by arc-length and of the shock speed being the arithmetic average of the left and right characteristic speeds.

\begin{Prop}
Assume the hypotheses of Lemma \eqref{shock-param-lemma}  and assume the $k$-th characteristic field is genuinely nonlinear with $r_k\ewl(\lambda_k)>0$. Then the curve $\epsilon \mapsto w_k(\epsilon)$ defined in \eqref{curve_state-space}, is $C^2$ regular and is parameterized by arc-length. Moreover, the shock speed satisfies
\beq \label{speed_shock_approx}
s_k(\epsilon) = \frac{\lambda_k(w_l) - \lambda_k(\epsilon)}{2} + O(\epsilon^2). 
\eeq
\end{Prop}
\Proof
Equation \eqref{speed_shock_approx} follows by the proof of Theorem 17.16 in \cite{Smoller}. Note that $r_k\ewl(\lambda_k)$ cancels out in \eqref{speed_shock_approx}, so that the normalization $r_k\ewl(\lambda_k)=1$ in \cite{Smoller} is irrelevant for the result here.

The rest of the corollary is immediate from Theorem \ref{Thm_simple-waves} and Lemma \ref{shock-param-lemma}. In particular, arc-length parameterization follows since $\langle r_k,r_k\rangle_\M(w)=1$ for each $w\in\M$ and since the rarefaction curves satisfy $\tfrac{d}{d\epsilon}{\hat{w}}_k(\epsilon)=r_k(w_k(\epsilon))$, by Theorem \ref{Thm_simple-waves}.
\QED

\subsection{Contact Discontinuities} \label{sec_Thm_mfd_contact-disc}

We now solve the Riemann problem \eqref{system_cons_laws_manifold} and \eqref{Riemann_problem} when the $k$-th characteristic field is linearly degenerate, c.f. Definition \ref{linearly degenerate}, within the class of contact discontinuities. We call a function $w(x,t)$ a \emph{contact discontinuity} if it is a weak solution of the Riemann problem \eqref{system_cons_laws_manifold} - \eqref{Riemann_problem}, discontinuous across the straight line $\{(st,t):t\geq0\}$, such that $s$ coincides with the characteristic speed on \emph{both} sides of that line. 

\begin{Thm}  \label{contact-discontinuity_Thm}
Let the $k$-th characteristic field be linearly degenerate in the sense of Definition \ref{linearly degenerate}. Then, given some $w_l \in \M$, there exists a one-parameter family of states, $w(\epsilon)$, for $\epsilon \in I$, for some interval $I$ containing $0$, such that for each $\epsilon \in I$, there exists a contact discontinuity solving the Riemann problem \eqref{system_cons_laws_manifold} - \eqref{Riemann_problem} for $w_l=w(0)$ and $w_r=w(\epsilon)$.        If \eqref{system_cons_laws_manifold} is strictly hyperbolic in the sense of Definition \ref{def_hyperbolicity}, then this one-parameter family is unique. 
\end{Thm}           

\Proof
Given some $w_l \in \M$, define the curve $\epsilon \mapsto w(\epsilon)$, for $\epsilon$ sufficiently close to $0$, as the solution of the ODE
\beq \label{contact_discont_curve}
\frac{d w}{d \epsilon} = r_k\big(w(\epsilon)\big),
\eeq
for initial data $w(0)=w_l$. The assumption of linear degeneracy, that is, $r_k\ew\big(\lambda_k\big) =0$ for all $w \in \M$, implies that $\lambda_k$ is constant along the curve $\epsilon \mapsto w(\epsilon)$. That is,
\beq\nn
\lambda_k\big(w(\epsilon)\big) = \lambda_k\big(w_l\big),
\eeq
for all $\epsilon \in I$, where $I$ denotes some open interval containing $0$. 

Now, given some $\epsilon \in I$, we define the function
\beq \label{contact_discontinuity}
v(x,t) \equiv \begin{cases} w_l, & x< t\, \lambda_k(w_l), \\ w(\epsilon), & x> t\, \lambda_k(w_l) .    \end{cases}
\eeq
Clearly, $v$ satisfies the initial data \eqref{Riemann_problem} for $w_r=w(\epsilon)$, and setting $s=\lambda_k(w_l)$, we find that
\beq\nn
\frac{d}{d\epsilon} \left( f\big(w(\epsilon)\big) - s g\big(w(\epsilon)\big) \right) = df\ew\big(r_k\big) - s dg\ew\big(r_k\big),
\eeq
which vanishes by the eigenvalue problem \eqref{eigenvalue-problem_right_mfd}. We conclude that
\beq\nn
 f\big(w(\epsilon)\big) - s g\big(w(\epsilon)\big)  =  f\big(w_l\big) - s g\big(w_l\big) ,
\eeq
for all $\epsilon \in I$, which are in fact the RH conditions, \eqref{RHconditions}. Therefore, $v$ satisfies the RH conditions and is a weak solution of \eqref{system_cons_laws_manifold}, according to Proposition \ref{weak-sol<=>RH-condition}. We conclude that $w_l$ and $w(\epsilon)$ are connected by a contact discontinuity. Uniqueness of the family of states follows from \eqref{contact_discont_curve} together with the uniqueness of the vectorfield $r_k$ by strict hyperbolicity.
\QED

\subsection{Proof of Theorem \ref{Thm_manifold}} \label{sec_Thm_mfd_proof}

Choose coordinates $y^\mu$ in some neighborhood $\U$ of $w_l$ in $\M$. We now follow the arguments in the proof of Theorem 17.18 in \cite{Smoller}. By the results from sections \ref{sec_Thm_mfd_rarefaction-waves} - \ref{sec_Thm_mfd_contact-disc}, there exists a neighborhood $\U \subset \M$ of $w_l$ and for each $k=1,..,n$ there exists a one-parameter family of maps 
\beq \nn
T^k_{\epsilon_k} : \U \rightarrow \M ,
\eeq
with the defining property that any $w\in\U$ can be joined to $T^k_{\epsilon_k} w$ by either a $k$-shock, a $k$-rarefaction wave or a contact discontinuity, depending on whether the $k$-th characteristic field is genuinely non-linear or linearly degenerate.  
Set $\epsilon\equiv (\epsilon_1,...,\epsilon_n)$ and define the composition $T(\epsilon) \equiv T^n_{\epsilon_n}...T^1_{\epsilon_1}$. To complete the prove it remains to show that there exists exactly one $\epsilon$ such that $w_r = T(\epsilon) w_l$. This follows by the Inverse Function Theorem. In more detail, defining 
\beq\nn
F(\epsilon_1,..,\epsilon_n) \equiv T^n_{\epsilon_n}...T^1_{\epsilon_1}w_l - w_l,
\eeq
it follows that $F(0,...,0) = 0$ and, by construction of the wave curves (see \eqref{contact_discont_curve} and Lemma \ref{shock-param-lemma}), that              
\beq\nn
\frac{\partial F}{\partial \epsilon_i}(0,...,0) = r_i(w_l).
\eeq
Now, since the $r_k$'s are linearly independent by the assumption of hyperbolicity, the Inverse Function Theorem implies the existence of a unique $\epsilon$ such that $w_r = T(\epsilon) w_l$. The result is independent of the choice of coordinates, since the intermediate states as well as the wave curves connecting them are independent of coordinates. This completes the proof of Theorem \ref{Thm_manifold}. \hfill $\Box$

\section{The Cauchy Problem and Glimm's Scheme}  \label{sec_Glimm}

In this section we extend Glimm's method for proving global existence of the Cauchy problem with initial data of bounded total variation to our framework of a manifold of states and prove Theorem \ref{Glimm_Thm_Intro}. 

To begin, we introduce the canonical distance function ${d}_\M$, which turns $\M$ into a metric space. 
That is, $d_\M : \M \times \M \rightarrow [0,\infty)$ is defined as        
\beq \label{distancefunction}
d_\M (p,q) = \inf_{\gamma} \int_a^b \sqrt{\langle \dot{\gamma}(s),\dot{\gamma}(s)\rangle_\M} ds,
\eeq
where $\langle \cdot,\cdot\rangle_\M$ denotes the metric tensor on $\M$ and the infimum is taken over all continuous piecewise differentiable curves $\gamma$ with $\gamma(a)=p$ and $\gamma(b)=q$.

Given a curve $w: [a,b] \rightarrow \M$, we now define the total variation of $w$ as
\beq \label{totalvariation_1step}
T.V.\big|_a^b(w) \equiv \sup_{\mathcal{P}} \sum_{i=0}^{n_p-1} {d}_\M\big( w(x_{i+1}), w(x_{i}) \big)
\eeq
where $\mathcal{P}$ denotes the set of partitions of the interval $[a,b]$ and $\{x_0,...,x_{n_p}\} \in \mathcal{P}$ is a partition. For a curve $w: \R \rightarrow \M$, we define its \emph{total variation} as
\beq \label{totalvariation}
T.V.(w) \equiv \lim_{k\rightarrow \infty} T.V.\big|^{\ k}_{-k}\big(w|_{[-k,k]}\big) .
\eeq
Given a constant state $\bar{w} \in \M$, we define 
\beq \label{supnorm}
d_\infty(\bar{w},w) \equiv \sup_{x\in \R} d_\M\big(w(x),\bar{w}\big),
\eeq               
which generalizes the sup-norm on states in $\R^n$ to our framework and which gives us a convenient notion of distance, sufficient to prove convergence of the Glimm scheme below. The main result of this section is the generalization of Glimm's famous Theorem \cite{Glimm}, recorded in Theorem \ref{Glimm_Thm_Intro}, which we now recall.

\begin{Thm} \label{Glimm_Thm}
Assume the system \eqref{system_cons_laws_manifold} is strictly hyperbolic and each characteristic field is genuinely non-linear or linearly degenerate in some neighborhood of some point $\bar{w}\in\M$. Given some curve                
\beq \nn
w_0: \R \rightarrow \M
\eeq 
such that $T.V.(w_0)$ and $d_\infty(w_0,\bar{w})$ are sufficiently small, then there exists a weak solution $w(x,t)\in\M$ of \eqref{system_cons_laws_manifold} for all $x\in \R$ and all $t\geq 0$ with initial data $w_0$, and there exists a constant $C>0$ such that     
\begin{eqnarray}\nn
T.V.\big(w(\cdot,t)\big) + d_\infty\big(w(\cdot,t),\bar{w}\big) &\leq & C\,\big( d_\infty(w_0,\bar{w}) + T.V.(w_0)  \big), \hspace{.3cm} \forall \ t\geq 0,  \cr
\int_{-\infty}^\infty d_\M\big( w(x,t_2),w(x,t_1) \big) dx &\leq & C\, |t_2-t_1| \ T.V.(w_0).
\end{eqnarray} 
\end{Thm}

The remainder of this section is devoted to the proof of Theorem \ref{Glimm_Thm}. Throughout this section, we assume \eqref{system_cons_laws_manifold} to be strictly hyperbolic and each characteristic field to be either genuinely non-linear or linearly degenerate.

\subsection{Interaction Estimates}

We begin the proof of Theorem \ref{Glimm_Thm} by deriving interaction estimates. For this, assume we are given three constant states, $w_l$, $w_m$ and $w_r$, sufficiently close to $\bar{w}$ such that their mutual Riemann problems are well-posed, c.f. Theorem \ref{Thm_manifold}. Assume further that 
\begin{eqnarray} \label{interactions}
w_r &=& T(\epsilon) w_l \ \equiv \ T^n_{\epsilon_n}...T^1_{\epsilon_1} w_l, \cr
w_m &=& T(\gamma) w_l \ \equiv \ T^n_{\gamma_n}...T^1_{\gamma_1} w_l, \cr
w_r &=& T(\delta) w_m \ \equiv \ T^n_{\delta_n}...T^1_{\delta_1} w_m ,
\end{eqnarray}
with $T$ denoting the operator introduced in section \ref{sec_Thm_mfd_proof}. We call the curve on $\M$ connecting $w_l$ with $w_r=T(\epsilon)w_l$ the $\epsilon$-wave, setting $\epsilon = (\epsilon_1,...,\epsilon_n)$. The curve connecting the state $T^j_{\epsilon_j}...T^1_{\epsilon_1} w_l$ with $T^{j-1}_{\epsilon_{j-1}}...T^1_{\epsilon_1} w_l$ is referred to as the $\epsilon_j$-wave, or simply as the $j$-wave, and we call $|\epsilon_j|$ the strength of the $j$-wave. 

\begin{Prop} \label{interactionestimate_Prop}
If $w_l$, $w_m$ and $w_r$ are sufficiently close to a constant state $\bar{w}\in\M$, then    
\beq \label{interactionestimate_1}
\epsilon_i = \gamma_i + \delta_i + O(|\gamma|\, |\delta|) ,
\eeq
where $|\gamma|\equiv \max \{|\gamma_i|:i=1,...,n\}$. Moreover, if there exists a coordinate system of Riemann invariants $\omega_j$ near $\bar{w}$, then            
\beq \label{interactionestimate_2}
\epsilon_i = \gamma_i + \delta_i + O\Big(\big(|\gamma| + |\delta|  \big)^3\Big) .
\eeq
\end{Prop}

\Proof
Choose local coordinates $y^\mu$ around $\bar{w}$ such that $w_l$, $w_m$ and $w_r$ are contained in the coordinate neighborhood. By \eqref{interactions}, $y_r \equiv y(w_r)$ is a $C^2$ function with respect to $\epsilon=(\epsilon_1,..,\epsilon_n) \in \R^n$. Taylor expanding $y_r$ around $\epsilon=0$, (keeping in mind that second order derivatives do in general not commute), gives
\beq \label{interactionestimates_Prop_eqn0}
y^{\, \mu}_r - y^{\, \mu}_l  \
= \ \sum_{j=1}^n \frac{\partial y^{\, \mu}_r}{\partial\epsilon_j}\bigg\rvert_0 \epsilon_j 
+  \sum_{j\leq k} \epsilon_j \epsilon_k  \left(1-\frac12\delta_{jk}\right) \frac{\partial^2 y^{\, \mu}_r}{\partial\epsilon_j\partial\epsilon_k}\bigg\rvert_0      + O(|\epsilon|^3)   
\eeq
where $\delta_{jk}$ denotes the Kronecker symbol. For genuinely non-linear fields, from \eqref{shock-param-lemma_curves} of Lemma \ref{shock-param-lemma}, we have $\frac{\partial y^{\, \mu}_r}{\partial\epsilon_j}\big\rvert_0 = r^\mu_j(w_l)$ and thus, for $j<k$,
\beq \nn
\frac{\partial^2 y^{\, \mu}_r}{\partial\epsilon_j\partial\epsilon_k}\bigg\rvert_0   
=  \frac{\partial r^\mu_k\big(T^{k-1}_{\epsilon_{k-1}}...T^{1}_{\epsilon_{1}}y_l\big)}{\partial \epsilon_j} \bigg\rvert_0  
=  r^\nu_j \partial_\nu (r^\mu_k)  \Big\rvert_{y_l} 
\equiv r_j(r^\mu_k)  \big\rvert_{y_l} ,
\eeq
where $\partial_\nu$ refers to partial differentiation in coordinates $y$ on $\M$, not to differentiation in parameter space $\epsilon=(\epsilon_1,...,\epsilon_n)$. For $j=k$, \eqref{shock-param-lemma_curves} implies that
\begin{eqnarray} \nn
\frac{\partial^2 y^{\, \mu}_r}{\partial\epsilon_k^2}\bigg\rvert_0   
&=&  \nabla_{\frac{\partial}{\partial \epsilon_k}}\frac{\partial y^{\, \mu}_r}{\partial\epsilon_k}\bigg\rvert_0  -  \Gamma^\mu_{\ \nu\sigma} \frac{\partial y^{\, \nu}_r}{\partial\epsilon_k}\frac{\partial y^{\,\sigma}_r}{\partial\epsilon_k}\bigg\rvert_0    \cr
&=&    \nabla_{r_k} r^{\,\mu}_k\Big\rvert_{y_l}  -  \Gamma^\mu_{\ \nu\sigma} r^{\, \nu}_k r^{\,\sigma}_k\Big\rvert_{y_l} \cr
&=&  r^\nu_k \partial_\nu (r^\mu_k)  \Big\rvert_{y_l}  
\ \equiv \ \  r_k (r^\mu_k)  \big\rvert_{y_l},
\end{eqnarray}
where $\Gamma^\mu_{\ \nu\sigma}$ is the Christoffel symbol of the covariant derivative $\nabla$. By \eqref{contact_discont_curve}, the above equations also hold true for a contact discontinuity of a linearly degenerate field. Substituting the above identities into \eqref{interactionestimates_Prop_eqn0} yields 
\beq \label{interactionestimates_Prop_eqn1}
y^{\, \mu}_r - y^{\, \mu}_l 
= \sum_{j=1}^n r^{\,\mu}_j\Big\rvert_{y_l} \epsilon_j 
+ \sum_{j\leq k}   \left(1-\frac12\delta_{jk}\right) r_k\big( r^{\,\mu}_j \big)\Big\rvert_{y_l} \epsilon_j \epsilon_k          
+ O(|\epsilon|^3).
\eeq

Likewise, we find for $y_m=T(\gamma)y_l$ the expression
\beq \label{interactionestimates_Prop_eqn1b}
y^{\, \mu}_m   -  y^{\, \mu}_l
= \sum_{j=1}^n r^{\,\mu}_j\Big\rvert_{y_l} \gamma_j 
+ \sum_{j\leq k}   \left(1-\frac12\delta_{jk}\right) r_k\big( r^{\,\mu}_j \big)\Big\rvert_{y_l} \gamma_j \gamma_k          
+ O(|\gamma|^3),
\eeq
while Taylor expanding $y_r=T(\delta)y_m$ at $y_m$ gives
\beq \label{interactionestimates_Prop_eqn1c}
y^{\, \mu}_r - y^{\, \mu}_m 
= \sum_{j=1}^n r^{\,\mu}_j\Big\rvert_{y_m} \delta_j 
+ \sum_{j\leq k}   \left(1-\frac12\delta_{jk}\right) r_k\big( r^{\,\mu}_j \big)\Big\rvert_{y_m} \delta_j \delta_k          
+ O(|\delta|^3).
\eeq
To compare \eqref{interactionestimates_Prop_eqn1} - \eqref{interactionestimates_Prop_eqn1c}, we now derive an expression for \eqref{interactionestimates_Prop_eqn1c} which is centered at $y_l$. For this, we Taylor expand $r_j|_{y_m}$ and $r_k(r^{\,\mu}_j)|_{y_m}$ at $y_l$ and apply \eqref{shock-param-lemma_curves}, which yields
\begin{eqnarray}\nn
r^{\,\mu}_j\Big\rvert_{y_m} &=&  r^{\,\mu}_j\Big\rvert_{y_l} + \sum_{i=1}^n r_i\big( r^{\,\mu}_j\big)\Big\rvert_{y_l} \gamma_i + O(|\gamma|^2),  \cr
r_k \big(  r^{\,\mu}_j\big)\Big\rvert_{y_m} &=& r_k \big(  r^{\,\mu}_j\big)\Big\rvert_{y_l} + O(|\gamma|) . 
\end{eqnarray}
Substituting the previous two identities into \eqref{interactionestimates_Prop_eqn1c} yields
\begin{eqnarray}\label{interactionestimates_Prop_eqn2}
y^{\, \mu}_r - y^{\, \mu}_m 
&=& \sum_{j=1}^n r^{\,\mu}_j\Big\rvert_{y_l} \delta_j + \sum_{i,j=1}^n r_i\big( r^{\,\mu}_j\big)\Big\rvert_{y_l} \gamma_i \delta_j   \cr
&+&  \sum_{j\leq k}   \left(1-\frac12\delta_{jk}\right) r_k \big( r^{\,\mu}_j \big)\Big\rvert_{y_l} \delta_j \delta_k          
+ O\left(|\delta|^3+|\gamma|\, |\delta|^2\right)
\end{eqnarray}

To complete the proof, we add \eqref{interactionestimates_Prop_eqn2} to \eqref{interactionestimates_Prop_eqn1b} and equate the resulting expression to the right hand side of \eqref{interactionestimates_Prop_eqn1}, which gives
\begin{eqnarray} \label{interactionestimates_Prop_eqn3}
\sum_{i=1}^n \left( \epsilon_i - \gamma_i - \delta_i \right)r^{\,\mu}_i 
&=&   \sum_{j\leq k}   \left(1-\frac12\delta_{jk}\right) r_k\big( r^{\,\mu}_j \big)\Big\rvert_{y_l} \left(\gamma_j \gamma_k + \gamma_j \gamma_k - \epsilon_j \epsilon_k \right) \ \ \  \  \cr 
&+&  \sum_{i,j=1}^n r_i\big( r^{\,\mu}_j\big)\Big\rvert_{y_l} \gamma_i \delta_j  + O\Big(\big(|\gamma| + |\delta|  \big)^3+|\epsilon|^3\Big).
\end{eqnarray}
Using now that $\epsilon_i = O(|\gamma|+|\delta|)$, since $\epsilon$ vanishes when $\gamma$ and $\delta$ are zero, we conclude from \eqref{interactionestimates_Prop_eqn3} that
\beq \label{interactionestimates_Prop_eqn4}
\epsilon_j = \gamma_j + \delta_j + O\left( (|\gamma|+|\delta|)^2 \right).
\eeq
Substituting \eqref{interactionestimates_Prop_eqn4} into \eqref{interactionestimates_Prop_eqn3}, a straightforward computation finally yields
\beq \nn
\sum_{i=1}^n \left( \epsilon_i - \gamma_i - \delta_i \right)r^{\,\mu}_i 
= \sum_{j<i} \gamma_i \delta_j \left(r_i\big(r^{\,\mu}_j\big) - r_j\big(r^{\,\mu}_i\big) \right) + O\Big(\big(|\gamma| + |\delta|  \big)^3\Big),
\eeq 
from which \eqref{interactionestimate_1} is immediate. Moreover, since the existence of a coordinate system of Riemann invariants implies the Lie brackets of the $r_k$'s in the previous equation to vanish, \eqref{interactionestimate_2} follows. 
\QED

From Proposition \ref{interactionestimate_Prop} one obtains the main interaction estimate, recorded in the following theorem. Before we state the theorem, consider the interacting waves in \eqref{interactions}. We say the $j$-wave in $\gamma$ and the $k$-wave in $\delta$ \emph{approach} if either $j>k$ or if $j=k$ and at least one of the waves is a shock. (We often say $\gamma$ in place of $\gamma$ wave and so on.) That is, two waves approach if the wave on the left is faster than the wave on the right. Note that neither rarefaction waves nor the contact discontinuities constructed in Theorem \eqref{contact-discontinuity_Thm} interact, if they belong to the same characteristic family, since the head of the left wave travels with the speed of the tail of the right wave.  We now define
\beq \nn
D(\gamma,\delta) \equiv \sum |\gamma_i||\delta_j|,
\eeq
where the sum is over all pairs for which the $i$ wave in $\gamma$ and the $j$-wave in $\delta$ approach.

\begin{Thm} \label{interactionestimate_Thm}
If $w_l$, $w_m$ and $w_r$ are sufficiently close to $\bar{w}$, then \footnote{The $O(1)$ in \eqref{interactionestimate_main} should be understood as a constant weight to each term in $D$ and not as a constant.}
\beq \label{interactionestimate_main}
\epsilon_i = \gamma_i + \delta_i + D(\gamma,\delta)O(1) \ \ \ \ \ \ \text{as} \ \ \ |\gamma|+|\delta| \rightarrow 0, 
\eeq
and if \eqref{interactionestimate_2} hold, then     
\beq \nn
\epsilon_i = \gamma_i + \delta_i + D(\gamma,\delta)O\big(|\gamma| + |\delta|  \big) \ \ \ \ \ \ \text{as} \ \ \ |\gamma|+|\delta| \rightarrow 0.
\eeq
\end{Thm}

\Proof   
The proof of Theorem 19.2 in \cite{Smoller} can be taken word by word.
\QED

\subsection{The Random Choice Method}

We now introduce Glimm's random choice method for constructing approximate solutions, c.f. \cite{Glimm,Smoller}. Adapting the scheme to our framework is straightforward and we include it only for completeness. To begin, divide the line $\{t= k \Delta t\}$ into the segments lying between the points $m\Delta x$, $m\in\Z$ with $k+m$ being an even number and $\Delta x$, $\Delta t>0$. For $k \in \mathbb{N}$, this defines a grid over $\R\times[0,\infty)$. We let the ratio $\frac{\Delta x}{\Delta t} \equiv c$ be fixed and assume the stability condition
\beq \label{nointeractions}
\frac{\Delta x}{\Delta t} \, >\, \sup \{|\lambda_j(w)| : w\in \hat{\U}, 1\leq j \leq n  \},
\eeq
to ensure that waves from different grid cells do not interact within a single time step. Here $\hat{\U}$ denotes the set containing $\U$ such that the intermediate states of the solution to the Riemann problem $(w_l,w_r)$ lie in $\hat{\U}$ for all $w_l$, $w_r \in \U$. Note that, by continuous dependence of the wave curves on their emanating states $w_l$, $\hat{\U}$ exist for $\U$ sufficiently small. 

Now, choosing for each $k\in\mathbb{N}$ some $\theta_k \in [-1,1]$ randomly, the \emph{mesh points} are defined as
\beq \nn
a_{m,k} \equiv (m\Delta x + \theta_k \Delta x, k \Delta t) ,
\eeq
for all $m \in \mathbb{Z}$ and $k \in \mathbb{N}$ with $k+m$ even. Assuming now some constant values $w^{k-1}_{m-1} \in \M$ and $w_{m+1}^{k-1} \in \M$ are given, and let $v_{m,k}(x,t)\in \M$ denote the solution of the Riemann problem 
\beq \nn
g(v)_t + f(v)_x =0
\eeq 
with initial data at $t=(k-1)\Delta t$ given by
\beq \nn
v\big(x, (k-1)\Delta t\big) = \begin{cases} w_{m-1}^{k-1}, \ \ \ \ (m-1)\Delta x \leq x \leq m \Delta x ,\cr w_{m+1}^{k-1}, \ \ \ \ m\Delta x \leq x \leq (m+1) \Delta x .\end{cases}
\eeq
We then introduce the constant state
\beq \nonumber
w_{m}^k \equiv v_{m,k}(a_{m,k}),
\eeq
for $m\in \Z$ with $m+k$ even. 

This defines Glimm's random choice scheme. Given constant states $w_m^0 \in \M$, each state $w_m^0$ being associated to the interval $m \Delta x \leq x \leq (m+2) \Delta x$ at $t=0$, the above scheme defines the values $w_m^k \in \M$ for all $k\in\mathbb{N}$ and $m\in \Z$, provided all such values stay in the neighborhood $\U$ where the above Riemann problem is well-posed. We denote the approximate solution obtain by the above scheme with 
\beq \label{def_approximates_Glimm-scheme}
w_{\theta,\Delta x}(x,t) \equiv v_{m,k}(x,t),   
\eeq 
for $x \in [(m-1)\Delta x, (m+1)\Delta x]$ and $t\in [(k-1)\Delta t, k\Delta t]$, where $k+m$ is assumed even. In Theorem \ref{Thm_functionals}, we prove that $w_{\theta,\Delta x}$ is indeed defined for all $t\geq 0$, by showing that $w_{m}^k \in \U$ for all $k\in\mathbb{N}$ and $m\in \Z$ with $k+m$ even.

\subsection{The Glimm Functional and Total Variation Bounds}

As in \cite{Glimm,Smoller} we define \emph{mesh curves} as (unbounded) piecewise linear curves connecting mesh points by moving from ``west'' to either ``north'' or ``south'', that is, a curve $I$ connecting mesh points is a mesh curve if it is linear in between mesh points, for each $k\in \Z$ there exists exactly one $n\in \mathbb{N}$ such that $a_{k,n}\in I$ and, given some $a_{k,n} \in I$, either $a_{k+1,n+1}\in I$ or $a_{k+1,n-1}\in I$. 

Given a mesh curve $I$, we define $I^-$ as the set of mesh points in the $t\geq 0$ half-plane below $I$ and $I^+$ as the set of mesh points above $I$. Given two mesh curves $I$ and $J$, we say that $I>J$ if every point on $I$ is either on $J$ or lies in $J^+$. This gives a partial ordering on the set of mesh curves. We say $I$ is an immediate successor of $J$ if $I>J$ and each point but one of $I$ lies also on $J$.

Following \cite{Glimm}, we define the following functionals on the mesh curve, $J$,
\begin{eqnarray}
L(J) &\equiv& \sum \{|\alpha|: \alpha \ \text{crosses}\  J \}, \label{functional_linear} \\
Q(J) &\equiv& \sum \{|\alpha| |\beta|: \alpha,\beta \  \text{cross}\  J \ \text{and approach}\}, \label{functional_nonlinear}
\end{eqnarray}
where we refer to the wave $w_r=T(\alpha)w_l$, solving the Riemann problem for initial data $w_l$ and $w_r$, as the $\alpha$ wave. Moreover, assuming $\alpha$ is a $j$-wave and $\beta$ is a $k$-wave, both crossing the mesh curve $J$, we say that $\alpha$ and $\beta$ approach if either $j > k$ and $\alpha$ lies to the left of $\beta$, or $j<k$ and $\beta$ lies to the left of $\alpha$, or $j=k$ and $\alpha$ and $\beta$ are distinct waves (not connecting the same states) and at least one of them is a shock wave. The next lemma is crucial for proving bounds on the above functionals.

\begin{Lemma} \label{L_equiv_TV}
For the set of approximates $w_{\theta,\Delta x}$, defined in \eqref{def_approximates_Glimm-scheme}, $T.V.(\cdot)$ and $L(\cdot)$ are equivalent in the sense that there exists a constant $C>0$, independent of $\theta$ and $\Delta x$, such that for any mesh curve $I$ we have  
\beq \label{L_equiv_TV_inequality}
\frac1C \,L(I) \ \leq \ T.V.(w_{\theta,\Delta x}|_I) \ \leq \ C \,L(I).
\eeq
\end{Lemma}

\Proof
To begin, consider two states $w_l, w_r \in \M$ such that $w_r$ is connected to $w_l$ on the right by either a $j$-shock, $j$-rarefaction wave or a $j$ contact discontinuity. We denote the corresponding curve by $\alpha$ and assume arc-length parameterization $\epsilon \mapsto \alpha(\epsilon)$ as in Lemma \ref{shock-param-lemma}. Now, it follows from the definition of $d_\M$, \eqref{distancefunction}, that 
\beq \nn
d_\M(w_l,w_r) 
\ \leq \ \int_a^b \sqrt{\langle\dot{\alpha},\dot{\alpha}\rangle_{\M}}\, d\epsilon ,
\eeq
where by \eqref{curve_state-space} $a$ and $b$ are such that $a=0$ and $\alpha(b)=w_r$ for $\alpha$ being a rarefaction curve, while $b=0$ and $\alpha(a)=w_r$ if $\alpha$ is a shock curve. In fact, $|b-a|$ is the strength of the wave. Now, since arc-length parameterization implies $\langle\dot{\alpha},\dot{\alpha}\rangle_{\M}=1$, and since $w_r= T^j_{\alpha_j} w_l$, the above inequality reduces to
\beq \label{L_equiv_TV_techeqn1}
d_\M(w_l,w_r) 
\ \leq \ |b-a| \ = \ |\alpha_j|.
\eeq

Now, the total variation of $w\equiv w_{\theta,\Delta x}$ on some mesh curve $I$ is the sum over the mutual distances (with respect to $d_\M$) of all intermediate states corresponding to the waves crossing $I$. Thus, applying \eqref{L_equiv_TV_techeqn1} for all these intermediate states, with $w_l$ and $w_r$ playing the role of the individual intermediate states, immediately gives
\beq \label{L_equiv_TV_techeqn1b}
T.V.(w|_I) \leq L(I).
\eeq

To derive a lower bound on $T.V.(w)$, assume again two states $w_l, w_r \in \M$ such that $w_r$ is connected to $w_l$ on the right by a $j$-wave $\alpha_j$, that is, $w_r= T^j_{\alpha_j} w_l$. We again assume the $j$-wave to be parameterized by arc-length and, for simplicity, we assume $a=0$ and $b=\alpha_j$. We now write
\beq\nn
|\alpha_j| \, = \, \int_a^b \,d\epsilon \,=\, \frac1{|\dot{\gamma}_0(0)|_\M} \int_a^b |\dot{\gamma}_0(0)|_\M \, d\epsilon,
\eeq
where $\gamma_0$ denotes the unique geodesic curve\footnote{When the infimum of $d_\M$ in \eqref{distancefunction} is attained, then the corresponding curve is a so-called geodesic curve, which is determined through the metric tensor alone. For $p$ and $q$ sufficiently close, there always exist a unique geodesic curve connecting both points and minimizing \eqref{distancefunction}. Moreover, geodesic curves are one degree more regular than the metric, thus $d_\M$ has the same regularity as the metric.} 
with $\gamma_0(a)=w_l$ and $\gamma_0(b)=w_r$ and where $|\cdot|_\M^{\ 2}\equiv \langle \cdot,\cdot\rangle_\M $. Using that geodesic curves have a constant parameterization, i.e., $|\dot{\gamma}_0(0)|_\M=|\dot{\gamma}_0(\epsilon)|_\M $ for all $\epsilon\in[a,b]$, it follows that 
\beq \label{L_equiv_TV_techeqn3}
|\alpha_j| \ = \ \frac1{|\dot{\gamma}_0(0)|_\M} \int_a^b |\dot{\gamma}_0(\epsilon)|_\M \, d\epsilon 
\ =\ \frac1{|\dot{\gamma}_0(0)|_\M} d_\M (w_l,w_r).
\eeq
To show that \eqref{L_equiv_TV_techeqn3} is indeed valid, it remains to prove that $|\dot{\gamma}_0(0)|_\M>0$ for $b>0$ sufficiently small. For this, we denote $\alpha_j$ by $\alpha$ and Taylor expand $\alpha(b) - \gamma_0(b)$ at $0$, using that $\alpha(b) = w_r= \gamma_0(b)$ and using that, by definition, second order covariant derivatives of geodesic curves vanish ($\nabla_{\tfrac{\partial}{\partial\epsilon}} \dot{\gamma_0}=0$), which leads to             
\begin{eqnarray}\nn
\dot{\gamma}^\mu_0(0) - r^\mu_k(0)  
&=&b\, \frac{d^2\alpha^\mu}{d\epsilon^2}(0) + \, \frac{d^2\gamma^\mu_0 }{d\epsilon^2}(0)  + O(b^2) \cr
&=& b\, \nabla_{\tfrac{\partial}{\partial\epsilon}} \dot{\alpha}^\mu(0)    - b\, \Gamma^\mu_{\rho\sigma} \left( \dot{\alpha}^\rho \dot{\alpha}^\sigma - \dot{\gamma}^\rho_0 \dot{\gamma}^\sigma_0 \right)\big|_0   + O(b^2) \cr 
&=& b\, \nabla_{r_k} r_k^\mu\big|_{w_l}  - b\, \Gamma^\mu_{\rho\sigma} \left( r_k^\rho r_k^\sigma - \dot{\gamma}^\rho_0 \dot{\gamma}^\sigma_0 \right)\big|_{w_l}  + O(b^2)  ,
\end{eqnarray}
where we applied \eqref{shock-param-lemma_curves} of Lemma \ref{shock-param-lemma} for the last equality. Thus, taking the absolute value with respect to the metric yields
\beq \label{L_equiv_TV_techeqn2}
|\dot{\gamma}_0|_\M 
\geq  |r_k|_\M - b \left( \big|\nabla_{r_k} r_k\big|_\M + \big|\Gamma^\mu_{\rho\sigma} \left( r_k^\rho r_k^\sigma - \dot{\gamma}^\rho_0 \dot{\gamma}^\sigma_0 \right) \big|_\M + O(b) \right) ,
\eeq
\eqref{L_equiv_TV_techeqn2} implies that $|\dot{\gamma}_0|_\M>0$ for $b>0$ sufficiently small, since $\big|\nabla_{r_k} r_k\big|_\M$ is uniformly bounded in $\U$ (by the assumed $C^2$ regularity of $f$ and $g$) and since $|\dot{\gamma}_0(0)|_\M$ is bounded by the maximum of $|r_k(\epsilon)|_\M$ along $\alpha$, for geodesic curves minimizing arc-length.

Now, to obtain a bound on $L(I)$ in terms of the total variation of $w|_I$, we define 
\beq \nn
C^{-1} \equiv \inf \big\{ |\dot{\gamma}(0)|_\M \, \big| \, \gamma\,  \text{\small geodesic curve connecting two states in } \, \hat{\U} \big\},
\eeq 
where $\hat{\U}$ denotes the set containing $\U$ such that the intermediate states of the solution to the Riemann problem $(w_l,w_r)$ lie in $\hat{\U}$ for all $w_l$, $w_r \in \U$. 
By \eqref{L_equiv_TV_techeqn2} and continuous dependence, one can restrict $\U$ enough such that for any geodesic curve $\gamma$ we have a uniform upper and lower bound on $|\dot{\gamma}(0)|_\M$, (on the order of $|r_k|_\M$) which then implies that $\infty > C > 0$. Now, by \eqref{L_equiv_TV_techeqn3}, for the $C$ defined above, the lower bound for the total variation in \eqref{L_equiv_TV_inequality} is immediate.
\QED

The following theorem shows that the Glimm functional, $L(I) + k Q(I)$, is non-increasing for some constant $k>0$. This was the key insight in \cite{Glimm} to prove convergence of the scheme later on.    

\begin{Thm} \label{Thm_functionals}
Consider two mesh curves, $I$ and $J$, with $J>I$, and assume $I$ is in the domain of definition of $w_{\theta,\Delta x}$. If $L(I)$ is sufficiently small, then $J$ is in the domain of definition of $w_{\theta,\Delta x}$ and 
\begin{eqnarray}
\begin{split}\label{Glimm's functionals estimate}
 Q(J) & \leq & Q(I), \hspace{1.5cm} \\
 L(J) + k Q(J) & \leq & L(I) + k Q(I),
\end{split}
\end{eqnarray}
for a constant $k>0$ independent of $I$ and $J$. In particular, if $T.V.(w_0)$ is sufficiently small, then $w_{\theta,\Delta x}$ is defined in $\R\times [0,\infty)$ and across each mesh curve $\hat{I}$
\beq \label{totalvariation_bound}
T.V.(w_{\theta,\Delta x}|_{\hat{I}})   \leq C\ T.V.(w_0)
\eeq
for a constant $C>0$ independent of $J$, $\theta$ and $\Delta x$.
\end{Thm}

\Proof     
The proof of the estimates \eqref{Glimm's functionals estimate} for $J$ being an immediate successor of $I$ follows by the exact same words as in the proof of Theorem 19.5 in \cite{Smoller}. A straightforward computation then shows that 
\begin{eqnarray} \nn
L(J) + k Q(J)   \ \leq \
L(I) + k Q(I)   \ \leq \  L(I) + k L(I)^2  
\ \leq \   2 L(I),
\end{eqnarray}
provided we assume $L(I)< \frac{1}{k}$. By \eqref{L_equiv_TV_techeqn1b}, of the proof of Lemma \ref{L_equiv_TV}, we then obtain the inequality 
\beq \label{Thm_functionals_techeqn1}
T.V.(w_{\theta,\Delta x}|_{J})  \ \leq \ L(J) + k Q(J) \ \leq \ 2L(I),
\eeq 
from which we conclude that $w_{\theta,\Delta x}$ is defined for the immediate successor $J$ of $I$, as long that $L(I)$ is sufficiently small.  

The inequalities \eqref{Glimm's functionals estimate} and \eqref{Thm_functionals_techeqn1} now extend to general mesh curves $J>I$ by induction, since the constant $k$ comes from the $O(1)$ weight in Theorem \ref{interactionestimate_Thm} and thus only depends on the $r_k$'s and the Riemannian metric. Using then that \eqref{Thm_functionals_techeqn1} implies $w_{\theta,\Delta x}|_J$ to stay inside $\hat{\U}$, so that
\beq \label{Thm_functionals_techeqn2}
\sup\{|\lambda_j(w_{\theta,\Delta x}|_J)|, 1\leq j\leq n \} < \frac{\Delta x}{\Delta t},
\eeq
it follows that $w_{\theta,\Delta x}$ is defined for any $J$ of $I$, as long that $T.V.(I)$ is sufficiently small.  

To extend the domain of definition of $w_{\theta,\Delta x}$ to $t\geq 0$, we first introduce $0$ as the unique mesh curve closest to $\{t=0\}$, (connecting all points $a_{m,0}$ with $a_{m+1,1}$ for $m\in 2\mathbb{Z}$). We then obtain
\begin{eqnarray} \nn
L(J) + k Q(J) \ \leq \  L(0) + k Q(0)  
\ \leq \  L(0) + k L(0)^2  
\ \leq \   2 L(0),
\end{eqnarray}
provided we assume $L(0) < \frac{1}{k}$. Now Lemma \ref{L_equiv_TV} yields again 
\beq \label{Thm_functionals_techeqn1b}
T.V.(w_{\theta,\Delta x}|_{J})  \ \leq \ L(J) + k Q(J) \ \leq \ C\ T.V.(0),
\eeq 
which implies again that the wave speeds of $w_{\theta,\Delta x}$ are bounded by $\frac{\Delta x}{\Delta t}$, c.f. \eqref{Thm_functionals_techeqn2}, so that $w_{\theta,\Delta x}$ is defined for all $t\geq 0$. 

By \eqref{Thm_functionals_techeqn1}, to prove \eqref{totalvariation_bound}, it remains only to bound $T.V.(0)$ by $T.V.(w_0)$. For this, consider a single Riemann problem $(w_l,w_r)$ which is resolved through the wave $w_r=T(\epsilon)w_l$ and denote the intermediate states by $w^i\equiv T^i_{\epsilon_i}...T^1_{\epsilon_1}w_l$, $i=0,...,n$. With this notation we now define the constant
\beq \nn
c \equiv \sup \left\{ \frac{d_\M(w^{i+1},w^{i})}{d_\M(w_l,w_r)} \bigg|\, i\in \{0,...,n-1\}, \  w_l, w_r \in \U  \right\},
\eeq         
which is finite for $\U$ sufficiently small, since in the critical case when $d_\M(w_l,w_r)$ is smaller than $d_\M(w^{i+1},w^{i})$, we have that $d_\M(w^{i+1},w^{i})$ converges to $0$ faster than $d_\M(w_l,w_r)$, as $w_l \rightarrow w_r$. It follows that             
\beq \nn 
T.V.(0)\leq c\,n\,T.V.(w_0),
\eeq 
where we used that $T.V.(w_0)$ is larger than the total variation of the piece-wise constant approximate of $w_0$ which are used as the initial states in the Glimm scheme. Now \eqref{totalvariation_bound} immediately follows from \eqref{Thm_functionals_techeqn1b}. 
\QED

The convergence of the $w_{\theta,\Delta x}$ as $\Delta x \rightarrow 0$, is in fact a consequence of the estimates of the next corollary.

\begin{Corollary} \label{Cor_estimates_Glimm}
If $T.V.(w_0)$ and $d_\M(w_0,\bar{w})$ are both sufficiently small, where $\bar{w}\in\M$ is some constant state, then, for all $t\geq 0$,             
\begin{align}     
T.V.\big(w_{\theta,\Delta x}(\cdot,t)\big) + d_\infty \big(w_{\theta,\Delta x}(\cdot,t) ,\bar{w} \big)\  \leq\   C\, \Big( T.V.(w_0) + d_\infty(w_0,\bar{w}) \Big), \label{TV_estimate} \\ 
\int_{-\infty}^{\infty} d_\M\big( w_{\theta,\Delta x}(x,t_2),w_{\theta,\Delta x}(x,t_1)  \big) dx\ \leq\  C\, \big( |t_2-t_1| + \Delta t \big)\ T.V.(w_0), \label{Lipschitz_estimate}
\end{align}    
for $C>0$ denoting some constant independent of $\theta$, $\Delta x$, $k$, $\Delta t$, $t$, $t_1$ and $t_2$.
\end{Corollary}

\Proof
For ease of notation we subsequently write $w$ instead of $w_{\theta,\Delta x}$ and use $C$ as a universal constant. Since the waves in $w_{\theta,\Delta x}$ from different grid cells do not interact due to \eqref{nointeractions}, it follows that
\begin{eqnarray} \nn
d_\infty\big(w(\cdot,t),\bar{w}\big) &=& d_\infty\big(w(\cdot,k\Delta t),\bar{w}\big), \cr
T.V.\big(w(\cdot,t)\big) &=& T.V.\big(w(\cdot,k\Delta t)\big)
\end{eqnarray}
for all $t \in \big( (k-1)\Delta t,k\Delta t\big]$. It thus suffices to prove \eqref{TV_estimate} on the lines $t=k\Delta t$ only. Moreover, by the previous equation, the estimate 
\beq \label{Cor_estimates_Glimm_techeqn1b}
T.V.\big(w(\cdot,k\Delta t)\big) \leq   C\ T.V.(w_0)
\eeq 
is an immediate consequence of \eqref{totalvariation_bound} in Theorem \ref{Thm_functionals}. To derive the remaining bound on $d_\infty$ in \eqref{TV_estimate}, observe that
\beq \label{Cor_estimates_Glimm_techeqn1}
d_\infty\big(w(\cdot,k\Delta t), \bar{w}\big) \leq  d_\infty\big(w(\cdot,k\Delta t), w(x_0,k\Delta t) \big) 
+\ d_\M\big(w(x_0,k\Delta t), \bar{w}\big) 
\eeq
for any $x_0\in \R$. For the first term on the right hand side we find that
\begin{eqnarray} \nn
d_\infty\big(w(\cdot,k\Delta t), w(x_0,k\Delta t) \big) 
&\leq &  \lim_{l\rightarrow \infty} \sup _{x\in [-l,l]} d_\M \big(w(\cdot,k\Delta t), w(x_0,k\Delta t) \big)  \cr
&\leq &  \lim_{l\rightarrow \infty} T.V.\big|^{\ l}_{-l}\big( w(\cdot,k\Delta t)\big) \cr
&\equiv & T.V.\big(w(\cdot,k\Delta t)\big) \cr
&\leq &  C\ T.V.(w_0),
\end{eqnarray}
where $T.V.\big|^{\ l}_{-l}$ is defined in \eqref{totalvariation_1step}. Thus \eqref{Cor_estimates_Glimm_techeqn1} is bounded by
\beq \nn
d_\infty\big(w(\cdot,k\Delta t), \bar{w}\big) 
\leq  C\ T.V.(w_0)  +\ d_\M\big(w(x_0,k\Delta t), \bar{w}\big)
\eeq
for any $x_0 \in \R$. Thus, as in \cite{Smoller}, taking the limit of the previous inequality as $x_0 \rightarrow \infty$ and using that $T.V.\big(w(\cdot,k\Delta t) \big) \leq C\ T.V.(w_0) < \infty$ implies that $w(x_0,k\Delta t)$ converges to some point $p\in\M$ as $x_0 \rightarrow \infty$, we get
\begin{eqnarray} \nn 
d_\infty\big(w(\cdot,k\Delta t), \bar{w}\big) 
&\leq &  C\ T.V.(w_0)  + d_\M(p,\bar{w}) .
\end{eqnarray}
By construction of the $w_{\theta,\Delta x}$, it follows that 
\beq \nn
p=\lim_{x\rightarrow \infty} w(x,k\Delta t) = \lim_{x\rightarrow \infty} w_0(x) \ \ \ \ \  \forall k\in\mathbb{N},
\eeq 
so that the previous inequality yields         
\begin{eqnarray} \nn 
d_\infty\big(w(\cdot,k\Delta t), \bar{w}\big) 
&\leq &  C\ T.V.(w_0)  + d_\infty(w_0,\bar{w}) ,
\end{eqnarray}
which together with \eqref{Cor_estimates_Glimm_techeqn1b} proves \eqref{TV_estimate}.

To derive the $L^1$ Lipschitz estimate \eqref{Lipschitz_estimate}, we follow the proof of Corollary 19.8 in \cite{Smoller}, only replacing the Euclidean distance function on $w(x,t_1)$ and $w(x,t_2)$ by the canonical distance function $d_\M$. In more detail, assume $t_2 > t_1$ and set $t_0 = \sup\{t\leq t_1: t=k\Delta t, \ k \in \mathbb{N}\} $  and $S= \tfrac{(t_2-t_0)}{\Delta t} + 1$. Now, fix some $x\in \R$ and define $I=[x-S\Delta x, x+S\Delta x]$, then $w(x,t_1)$ and $w(x,t_2)$ are determined by the Cauchy data $\{w(y,t_0):\ y\in I\}$.  Thus,       
\begin{eqnarray}\nn
d_\M\big( w(x,t_2), w(x,t_1) \big) 
&\leq & \sup\big\{d_\M\big(w(y,t_0) , w(x,t_0)\big):\, y\in I\big\} \cr
&\leq & C \ T.V.\big( \big\{w(y,t_0) :\, y\in I\big\}  \big)  
\end{eqnarray}
and defining $I_m=[(m-S)\Delta x, (m+1+S)\Delta x]$, we obtain
\begin{align} \nn
\int_{-\infty}^\infty d_\M\big( w(x,t_2), w(x,t_1) \big) 
&=& \sum_{m\in\mathbb{Z}} \int_{m\Delta x}^{(m+1)\Delta x} d_\M\big( w(x,t_2), w(x,t_1) \big)dx \cr
&\leq &  C\Delta x\sum_{m\in\mathbb{Z}} T.V.\big( \big\{w(y,t_0) : y\in I_m\big\}  \big) \hspace{.6cm} \cr
&\leq &  C(1+S) \Delta t\, \frac{\Delta x}{\Delta t} \ T.V.\big( w_0\big) , \hspace{2cm}
\end{align}   
which proves \eqref{Lipschitz_estimate} since $S\Delta t \leq t_2 - t_1 + 2 \Delta t$ and $\tfrac{\Delta x}{\Delta t} $ is fixed by \eqref{nointeractions}. 
\QED

\subsection{Convergence}

We first show that on each line $\{t=const.\}$ the approximates $w_{\theta,\Delta x}(\cdot,t)$ converge to a state $w(\cdot,t)$ as $\Delta x \rightarrow 0$. This follows from an extension of Helly's Theorem \cite{Natanson} to our framework, recorded in the next lemma.

\begin{Lemma}\label{Helly's Thm}
Let $\mathcal{F} \equiv \big\{u: [a,b] \rightarrow \M\big\}$ be an infinite family of curves and assume there exists a constant state $\bar{w}\in\M$ and a constant $K>0$ such that
\beq \nn
d_{\infty}(u,\bar{w}) < K \hspace{1cm} \text{and} \hspace{1cm} T.V.(u) < K, \hspace{1cm} \forall\, u \in \mathcal{F}.
\eeq
Moreover, assume $K$ is small enough for all $u\in \mathcal{F}$ to take values inside some coordinate patch $\U$ in the sense that $\overline{\{ u(x): u \in \mathcal{F}, x \in [a,b]\}} \subset \mathcal{U}$, where the closure is taken with respect to $d_\M$.\footnote{This smallness assumption on $K$ is not required for Helly's Theorem in $\R^n$. Replacing this assumption by requiring $(\M,d_\M)$ to be complete suffices to prove convergence of a sequence $(w_n)$ on countable subsets of $[a,b]$, by using the Bolzano Weierstrass Theorem on $(\M,d_\M)$ and a diagonal choice process, however, it is currently not clear to us how to extend this convergence to the whole interval $[a,b]$.} 
Then there exists a sequence $\big(w_k\big)_{k\in\mathbb{N}} \subset \mathcal{F}$ which converges point-wise for each $x\in [a,b]$ to a curve $w:[a,b] \rightarrow \M$ satisfying $T.V.(w)<K$.
\end{Lemma}

\Proof
Choose coordinates $y^\mu : \U \rightarrow \R^n$ such that $y^\mu(\bar{w})=0$ and such that the absolute value of each component of $y^\mu(p)$ is bounded by $d_\M(p,\bar{w})$ for each $p\in \U$. We now define for $u\in \mathcal{F}$
\beq \nn
\pi_u(x) \equiv  T.V.\big|^x_a(u) \hspace{1cm} \text{and} \hspace{1cm} \Omega_u^\mu(x) \equiv \pi_u(x) - y^\mu\big(u(x)\big),
\eeq
where the last definition is to be understood component-wise for $\mu \in \{1,..,n\}$ and $T.V.|^x_a$ is defined in \eqref{totalvariation_1step}. For each $u \in \mathcal{F}$, by the above choice of coordinates, $\pi_u$ and $\Omega^\mu_u$ are real-valued \emph{increasing} functions in $x$. By assumption, $\pi$ satisfies the bound
\beq \nn
\sup_{x\in [a,b]} \big| \pi_u(x) \big| < K \hspace{1cm} \forall u\in \mathcal{F}
\eeq 
and, since $d_\M$ is equivalent to the Euclidean distance whenever $d_\M$ is restricted to a coordinate patch, $\Omega^\mu_u$ satisfies for all $\mu=1,...,n$ the bound
\beq \nn 
\sup_{x\in [a,b]} \big| \Omega^\mu_u(x) \big| < C\,K \hspace{1cm} \forall u\in \mathcal{F}, 
\eeq
for some constant $C$ independent of $u$.

According to the Lemma preceding Helly's Theorem in \cite{Natanson}, for an infinite family of real valued increasing functions defined on $[a,b]$ with a uniform upper bound, there exist a sequence within that family converging at each point $x\in[a,b]$. Thus, there exists a sequence $(\pi_k)_{k\in\mathbb{N}} \subset (\pi_u)_{u\in\mathcal{F}}$ and some function $\pi$ such that 
\beq\nn
\lim_{k\rightarrow \infty} \pi_k(x) = \pi(x) \hspace{1cm} \forall x\in [a,b],
\eeq
and, choosing to each $\pi_k$ the corresponding $u_k \in \mathcal{F}$, there exists a  subsequence of the $\Omega^\mu_{k} \equiv \Omega^\mu_{u_k}$ such that
\beq\nn
\lim_{j\rightarrow \infty} \Omega^\mu_{k_j}(x) = \Omega^\mu(x) \hspace{1cm} \forall x\in [a,b], \ \ \forall \mu =1,..,n,
\eeq
for some function $\Omega^\mu$. We conclude that $y^\mu_{k_j}(x) \equiv \pi_{k_j}(x) - \Omega^\mu_{k_j}(x)$ converges to the function $y_0^\mu(x) \equiv \pi(x) -  \Omega^\mu(x)$. Finally, since the closure of the images of all $w\in \mathcal{F}$ is assumed to be contained in the coordinate neighborhood $\U$, $y^{-1}$ is defined for $y_0^\mu(x)$ for all $x\in [a,b]$, and it follows that the sequence $w_{j}\equiv y^{-1}(y_{k_j}) \in \M$ converges pointwise for all $x\in[a,b]$ to $w \equiv y^{-1}(y_0^\mu)$ and that $T.V.(w)<K$. Since the convergence is point-wise, the result is independent of the choice of coordinates.  
\QED

By \eqref{TV_estimate}, Lemma \ref{Helly's Thm} applies to the family of $\big(w_{\theta,\Delta x}\big)_{\Delta x}$ and gives us for any \emph{bounded} interval in any of the lines $\{t=\text{const}\}$ a convergent subsequence and, by Cantor's diagonal choice process, we can extract for each $t=\text{const}$ a subsequence which converges for all $x\in  \{t=\text{const}\}$. By a further diagonal process, there exist a subsequence $(w_{\theta,\Delta x_l})_{l\in\mathbb{N}}$ of $\big(w_{\theta,\Delta x}\big)$ which converges on any of the lines $\{t=\text{const} \in \mathbb{Q}\}$ to some $w(x,t) \in \M$ as $k\rightarrow \infty$ (with $\Delta x_l \rightarrow 0$ as $l\rightarrow \infty$). To extend this convergence to all $t\geq0$, (as in \cite{Glimm}), set $w_l \equiv w_{\theta,\Delta x_l}$ and use \eqref{Lipschitz_estimate} to compute for $q\in \mathbb{Q}$, $q\geq 0$, 
\begin{eqnarray} \nn
&& \int_a^b d_\M\big( w_i(x,t),w_j(x,t) \big) dx \cr 
&& \hspace{.5cm} \leq  \int_a^b d_\M\big( w_i(x,t),w_i(x,q) \big) dx  +  \int_a^b d_\M\big( w_i(x,q),w_j(x,q) \big) dx  \cr && \hspace{.8cm} +  \int_a^b d_\M\big( w_j(x,q),w_j(x,t) \big) dx  \cr
&& \hspace{.5cm} \leq  C\, \big( |t-q| +  \Delta t_i + \Delta t_j \big)\ T.V.(w_0)  +  \int_a^b d_\M\big( w_i(x,q),w_j(x,q) \big) dx.
\end{eqnarray}
Now, given some $\epsilon >0$, we choose $q$ such that 
\beq\nn
C\,  |t-q| \ T.V.(w_0) < \frac{\epsilon}{2}
\eeq
and then choose $i$ and $j$ large enough for the remaining terms to be bounded by $\tfrac{\epsilon}{2}$. This shows that $d_\M\big( w_i(x,t),w_j(x,t) \big)$ converges to $0$ in $L^1_{\text{loc}}(\R)$ as~$i,j\rightarrow\infty$ for all $t\geq0$ which implies convergences point-wise almost everywhere in $x$. Thus $(w_l(x,t))_{l\in\mathbb{N}}$ is a Cauchy sequence in $(\U,d_\M)$ for almost every $x\in \R$ and, since $(\U,d_\M)$ is a complete metric space, (assuming without loss of generality that $\U$ is restricted sufficiently to be compact, completeness of $(\U,d_\M)$ follows by the Hopf-Rinow Theorem, c.f. \cite{doCamo}), it follows that $(w_l(x,t))_{l\in\mathbb{N}}$ converges to some $w(x,t)\in\M$ for almost every $(x,t)\in \R\times [0,\infty)$. In summary, as a consequence of \eqref{TV_estimate} and \eqref{Lipschitz_estimate}, we proved the following theorem:

\begin{Thm} \label{Thm_convergence}
Within the family of functions $\big(w_{\theta,\Delta x} \big)$ defined in \eqref{def_approximates_Glimm-scheme}, there exists a subsequence $(w_l)_{l\in\mathbb{N}}$ which converges almost everywhere in $\Omega\equiv \R\times \{t\geq 0\}$ to some $w$ as $l\rightarrow \infty$. (By definition, each $w_l \equiv w_{\theta,\Delta x_l}$ corresponds to a grid length $\Delta x_l$ such that $\Delta x_l \rightarrow 0$ as $l\rightarrow \infty$.) For the limit function $w$ there exist a constant $C>0$ such that for any $t\geq 0$ 
\beq \nn
T.V.\big(w(\cdot,t)\big) + d_\infty\big(w(\cdot,t),\bar{w}\big) \leq  C\,\big( d_\infty(w_0,\bar{w}) + T.V.(w_0)  \big).
\eeq
\end{Thm}

To complete the proof of Theorem \ref{Glimm_Thm}, it remains to show that the limit function $w$ is indeed a weak solution of \eqref{system_cons_laws_manifold}. To continue, we prove the following corollary.

\begin{Corollary}
Given a continuous function $f: \M \rightarrow \R^n$ and let $(w_l)_{l\in\mathbb{N}}$ be the convergent subsequence of $\big(w_{\theta,\Delta x} \big)$ shown to exist in Theorem \ref{Thm_convergence}, then $f(w_l) \rightarrow f(w)$ in $L^1_\text{loc}(\Omega)$ as $l\rightarrow \infty$. 
\end{Corollary}

\Proof
By continuity of $f$, since $(w_l)_{l\in\mathbb{N}}$ converges to $w$ point-wise almost everywhere, it follows that $(f(w_l))_{l\in\mathbb{N}}$ converges to $f(w)$ point-wise almost everywhere. Moreover, since $(w_l)_{l\in\mathbb{N}}$ is uniformly bounded with respect to $l$, $\big(f(w_l)\big)_{l\in\mathbb{N}}$ is uniformly bounded as well. Therefore, since $f(w_l) \in L^1_\text{loc}(\Omega)$ for all $l\in \mathbb{N}$, the Lebesque dominated convergence Theorem implies that $(f(w_l))_{l\in\mathbb{N}}$ converges to $f(w)$ in $L^1_\text{loc}(\Omega)$.
\QED

We continue with the following fundamental observation \cite{Glimm}. Given $\psi \in C^1_0(\Omega)$, define 
\beq \nn
J(\theta, \Delta x,\psi) \equiv \iint\limits_{t\geq0} \big(g(w_{\theta,\Delta x})\psi_t + f(w_{\theta,\Delta x})\psi_x \big)dx dt + \int_{-\infty}^\infty g(w_{0_{\Delta x}}) \psi dx ,
\eeq 
where $w_{0_{\Delta x}}(x)$ is some function converging to $w_0(x)$ for almost every $x\in \R$ such that $g(w_{0_{\Delta x}})$ converges to $g(w_0)$ in $L^1_\text{loc}(\R)$, c.f. \cite{Glimm,Smoller}. A function $w: \Omega \rightarrow \M$ is a weak solution of \eqref{system_cons_laws_manifold} if and only if $J(\theta, \Delta x,\psi)=0$ for all $\psi  \in C^1_0(\Omega)$, c.f. Definition \ref{weak_solution}. By \eqref{def_approximates_Glimm-scheme}, $w_{\theta,\Delta x}$ is an (exact) weak solution in each time strip $k\Delta t \leq t \leq (k+1) \Delta t$, from which we get that
\begin{eqnarray} \label{weak_form_functional}
J(\theta, \Delta x,\psi) 
&=& \sum\limits_{k=1}^\infty \int_{-\infty}^{\infty} \psi(x,k\Delta t)\, \big[ g( w_{\theta,\Delta x})\big]_k(x)\,  dx 
\end{eqnarray} 
where
\beq\nn
\big[ g( w_{\theta,\Delta x})\big]_k(x) \equiv 
g\big( w_{\theta,\Delta x}(x,k\Delta t +0) \big) - g\big( w_{\theta,\Delta x}(x,k\Delta t -0) \big) .
\eeq

From \eqref{weak_form_functional} we obtain the non-trivial result that within the sequence $w_k=w_{\theta,\Delta x_k}$ there exists a subsequence for which $J(\theta, \Delta x,\psi) \rightarrow 0$ as $\Delta x \rightarrow 0$ for almost all $\theta$. To formulate this final lemma, define the space of random variables
\beq \nn
\Phi \equiv  \big\{(\theta_k)_{k\in\mathbb{N}} : \ \theta_k \in [-1,1]\ \forall\, k\in\mathbb{N} \big\}.
\eeq 

\begin{Thm} \label{Lemma_weak-convergence}
There exists a set $N \subset \Phi$ of measure zero and a sequence $\Delta x_i \rightarrow 0$, as $i\rightarrow \infty$, such that for any $\theta \in \Phi \setminus N$ and any test function $\psi \in C^1_0(\Omega)$, we have $J(\theta, \Delta x_i,\psi) \rightarrow 0$ as $i\rightarrow \infty$.
\end{Thm}

\Proof
The proof follows along the line of argument given in between Lemma 19.12 and Theorem 19.14 in \cite{Smoller}, see also \cite{Glimm}, by replacing all functions $u_{\theta,\Delta x}$ by $g(w_{\theta,\Delta x})$ and using $J(\theta, \Delta x,\psi)$ as defined above.  However, to prove that the total variation bound \eqref{TV_estimate} implies the point-wise bound on $|J(\theta,\Delta x, \Phi)|$ in terms of $\Delta x$ and $\|\Phi\|_\infty$, stated in Lemma 19.12 in \cite{Smoller}, we also need to use the total variation bound of the next lemma, Lemma \ref{TV_of_g_lemma}, which applies since one can always restrict $\U$ enough for $\overline{\U}$ to be compact and to lie in some coordinate neighborhood $(y,\U')$. Finally, a $L^2$ estimate of $J(\cdot,, \Delta x_i,\psi)$ with respect to integration over the random variable $\theta$ yields convergence almost everywhere, c.f. Theorem 19.4 in \cite{Smoller}.
\QED

\begin{Lemma} \label{TV_of_g_lemma}
Let $g:\M \rightarrow \R^n$ be Lipschitz continuous and consider a curve $w:\R \rightarrow \U$, for $\U\subset \M$ being a compact set contained in a coordinate neighborhood $(y,\U')$. Then 
\beq \label{TV_of_g}
T.V.\big(g(w)\big) \leq C\,  T.V.(w),
\eeq
for some constant $C>0$ which depends only on the metric tensor and the Lipschitz constant of $g$ in coordinates $y$, on the domain $y(\U)$. 
\end{Lemma}
\Proof
Let $w_1, w_2 \in \U$. Choose some coordinates $y$ on $\U$ and consider $y_1\equiv y(w_1)$ and $y_2\equiv y(w_2) \in y(\U)$, we then find
\begin{eqnarray} \label{TV_of_g_techeqn1}
|g\circ y^{-1}(y_1)-g\circ y^{-1}(y_2)|_\text{eucl} &\leq & \text{Lip}(g\circ y^{-1})\, |y_1-y_2|_\text{eucl},
\end{eqnarray}
where $\text{Lip}(g\circ y^{-1})$ denotes the Lipschitz constant of $g$ over $y(\U)$ and ~$|\cdot|_\text{eucl}$ denotes the Euclidean square norm. It remains to show that $|y_1-y_2|_\text{eucl} \leq C d_\M(y_1,y_2)$ for some constant $C>0$. To prove this, using that $\overline{\U}$ is assumed compact, we observe that there exist a constant $C>0$ such that
\beq \label{metric_bound}
|v|_\M \geq C |v|_\text{eucl}, \ \ \ \ \forall v \in T_p\M, \ \ \ \forall p \in \U.
\eeq  
(Here $C$ can be taken to be the supremum of the largest eigenvalue of the metric tensor, which is symmetric by definition.) Now, we consider some curve $\gamma : [a,b] \rightarrow y(\U)$ which connects $y_1$ with $y_2$ and compute, using \eqref{metric_bound}, that
\begin{eqnarray}  \nn
\int_a^b |\dot{\gamma}|_\M ds\ \geq\ C \int_a^b |\dot{\gamma}|_\text{eucl} ds \ \geq \ C |y_1-y_2|_\text{eucl},
\end{eqnarray}
where the last inequality holds since a straight line is the shortest connection between two points in Euclidean geometry. Since the curve $\gamma$ is arbitrary, we conclude that
\beq\nn
d_\M(y_1,y_2) \geq C \,|y_1-y_2|_\text{eucl}.
\eeq
From the above inequality and the fact that $d_\M(y_1,y_2)=d_\M(w_1,w_2)$ is invariant, \eqref{TV_of_g_techeqn1} implies
\begin{eqnarray} \nn
|g(w_1)-g(w_2)|_\text{eucl} \leq C \, \text{Lip}(g\circ y^{-1})\, d_\M(w_1,w_2),
\end{eqnarray}
from which we conclude \eqref{TV_of_g}.
\QED

Lemma \ref{TV_of_g_lemma} completes the proof of Theorem \ref{Lemma_weak-convergence}, and the following corollary is immediate.

\begin{Corollary} \label{convergence_corollary}
If $T.V.(w_0)$ and $d_\infty(w_0,\bar{w})$ are sufficiently small, then there exist a set of measure zero $N\subset \Phi$ and a sequence $\Delta x_i \rightarrow 0$, as $i\rightarrow \infty$, such that for any $\theta \in \Phi \setminus N$, $w_\theta \equiv \lim_{i\rightarrow\infty} w_{\theta,\Delta x_i}$ is a weak solution of the Cauchy problem \eqref{system_cons_laws_manifold} with initial data $w_0$.
\end{Corollary}

Corollary \ref{convergence_corollary} together with Theorem \ref{Thm_convergence} complete the proof of Glimm's Theorem extended to our geometric framework, that is, Theorem \ref{Glimm_Thm}.

\section{Riemann Problems for Constrained Conservation Laws} \label{sec_Thm_constraint}

In this section we prove Theorem \ref{Thm_constrained-system} and Proposition \ref{Thm_parameterization}. 
 
\vspace{.3cm}\noindent \emph{Proof of Theorem \ref{Thm_constrained-system}.}
By \eqref{Thm_constrained-system_cond1}, that is, the assumption that $$\text{det}\,[DG(u),DC(u)] \neq 0,$$ we conclude that the linear mapping $DC(u) :\R^m \longrightarrow \R^{m-n}$ is surjective at $u_l$. This implies that the set $\M\equiv C^{-1}(0)$ defines a $n$-dimensional submanifold in $\R^m$,  c.f. \cite[Theorem 5.1]{Spivak}. In more detail, assuming that $u_l\in \hat{\U}$ with $C(u_l)=0$, then, since $C$ is assumed to have full rank at $u_l$, the implicit function theorem yields that there exists an open set $\Omega \subset \R^n$ and a differentiable function $\varphi_{u_l}: \Omega \rightarrow \R^m$ invertible on its image such that $\varphi_{u_l}$ parameterizes the solution space $C^{-1}(0)$ sufficiently close to $u_l$. Since the implicit function theorem yields for any point $u \in \hat{\U}$ with $C(u)=0$ such a function $\varphi_{u}$ parameterizing some open subset of $C^{-1}(0)$, we conclude that \eqref{Thm_constrained-system_cond1} implies the solution space $\M \equiv C^{-1}(0)$ to be a $n$-dimensional submanifold of $\R^m$, c.f. Appendix \ref{sec_preliminaries}. Thus, the constrained system of $n\times m$ conservation laws, \eqref{constrained_system_eqn} - \eqref{constrained_system_constrained}, is a system of the form \eqref{system_cons_laws_manifold}, that is, a $n\times n$ system of conservation laws with states in the manifold $\M$. 

In order to apply Theorem \ref{Thm_manifold}, it remains to show that the constrained system \eqref{constrained_system_eqn} - \eqref{constrained_system_constrained} is strictly hyperbolic in the sense of Definition \ref{def_hyperbolicity}. For this, observe that condition \eqref{Thm_constrained-system_cond3}, that is, $DC r_k=0$, implies that the eigenvectors $r_k(u)$ lie in the tangent-space of $\M$ whenever $u\in\M$, so that by our hypothesis \eqref{Thm_constrained-system_cond2} the resulting system is strictly hyperbolic. Now, Theorem \ref{Thm_manifold} applies and yields the claimed existence and uniqueness of solutions, which completes the proof. 
\hfill $\Box$ \vspace{.3cm}

We now prove Proposition \ref{Thm_parameterization}, which shows that the manifold $\M$ is locally described by the integral curves of the eigenvectors $r_k$, and therefore by the rarefaction curves, c.f. \eqref{simple_wave_def}. This is useful in practice, because it only involves solving linear systems but never requires solving the constraint $C(u)=0$ explicitly which could be difficult to accomplish for many applications. One can then obtain the shock curves by substituting the local parameterization of $\M= C^{-1}(0)$ around $u_l$ in the Hugoniot locus.    We prove the following more precise version of Proposition \ref{Thm_parameterization}.

\begin{Prop} \label{Thm_parameterization_detailed}
Assume \eqref{Thm_constrained-system_cond1} - \eqref{Thm_constrained-system_cond3} hold for all $u\in \hat{\U}$ for some $\hat{\U} \subset \R^m$.  
Then, locally, the solution space $C^{-1}(0)$ is identical to the image of the integral curves of the eigenvectors $r_k$. That is, for each $u_l\in\hat{\U}$ with $C(u_l)=0$, there exist some $\U\subset \hat{\U}$  with $u_l\in \U$ such that 
\beq \nn
C^{-1}(0)\cap \U = \left\{ u \in \hat{\U} :\ u = S^n_{\epsilon_n} ... S^1_{\epsilon_1} u_l   \right\}
\eeq 
where $S^k_{\epsilon_k}u' \in \R^m$ is the point which is reached by the integral curve of $r_k$  emanating in $u'$ at a parameter $\epsilon_k$, for $k\in \{1,...,n\}$. Moreover, the integral curves of $r_k$ define a local parameterization of $C^{-1}(0)$ if and only if there exists a coordinate system of Riemann invariants.
\end{Prop}

To clarify, a Riemann invariant $\omega_j$ is a scalar function which satisfies $r^{\,\nu}_i \partial_\nu \omega_j =0$, for $i,j \in \{1,...,n\}$ with $i\neq j$. 

\Proof
Let $u_l\in  C^{-1}(0)$. To begin, we solve the ODE \ref{simple_wave_def} simultaneously with the eigenvalue problem \eqref{Thm_constrained-system_cond2} - \eqref{Thm_constrained-system_cond3}, that is, we solve the system
\beq \label{Thm_parameterization_eqn1}
\frac{d u_k}{d \epsilon} = r_k\big(u_k(\epsilon)\big), \ \ \ \ \ \ \text{for} \ \ \ \ \ u_k(0)=u_l ,
\eeq 
together with
\beq \label{Thm_parameterization_eqn2}
\left( \lambda_k\, DG\, -\, DF \right) r_k = 0  \ \ \ \ \text{and}  \ \ \ \ 
DC\, r_k = 0 , 
\eeq
the later two equations being evaluated at $u_k(\epsilon)$. Assuming for the moment the solution $(\lambda_k,r_k)$ of \eqref{Thm_constrained-system_cond2} - \eqref{Thm_constrained-system_cond3} exist everywhere in $\hat{\U}\subset \R^m$ and not only for $u\in  \hat{\U}$ with $C(u_l)=0$, it follows that the above system is well-posed, which gives us the integral curves $\epsilon \mapsto u_k(\epsilon)$, with $\epsilon \in (-a,a)$ for some $a>0$.                     

By the condition $DC\, r_k = 0$, it follows that $r_k(u) \in T_{u}\M$ whenever $u\in \M$. We now change coordinates in $\R^m$ such that some open neighborhood of $u_l$ in $\M$ coincides with some open set in $\R^n \times \{0\}^{m-n}$. It follows that $T_{u}\M$, as a subspace of $\R^m$, is identical to  $\R^n \times \{0\}^{m-n}$ for each $u$ in that neighborhood. Thus, using $u_k(0)=u_l \in \M$, we conclude that the curve $\epsilon \mapsto u_k(\epsilon)$ lies in $\M$ for all $\epsilon$ sufficiently small, since the $r_k(u_k(\epsilon))$ are in $\R^n \times \{0\}^{m-n}$ for all $\epsilon$. (This shows that the above system is well-posed even when the solution $(\lambda_k,r_k)$ of \eqref{Thm_constrained-system_cond2} - \eqref{Thm_constrained-system_cond3} exists only for $u\in C^{-1}(0)$.)  

To prove that $\M$ is locally ``parameterized'' by the integral curves $u_k$ for $k=1,...,n$, observe first that the system \eqref{Thm_parameterization_eqn1} - \eqref{Thm_parameterization_eqn2} is well-posed for any $u_l \in \M$. Now, for $\epsilon_k \in (-a,a)$, $k=1,...,n$, we introduce the mapping $S^k_{\epsilon_k} :\M \rightarrow \M$ as $S^k_{\epsilon_k} u =u_k(\epsilon_k)$ where $u_k$ is the integral curve of $r_k$ (in the sense of \eqref{Thm_parameterization_eqn1} - \eqref{Thm_parameterization_eqn2}) with initial data $u_k(0)=u$. Given $u'\in \M$ sufficiently close to $u_l$, it thus remains to show that there exist $(\epsilon_1,...,\epsilon_n)$ such that 
\beq \label{Thm_parameterization_eqn3}
u' = S^n_{\epsilon_n} ... S^1_{\epsilon_1} u_l.
\eeq
As in the proof of Theorem \ref{Thm_constrained-system}, this follows by the Inverse Function Theorem and the linear independence of the $r_k$'s. 

This proves that $\M \equiv C^{-1}(0)$ is locally described by the integral curves of $r_k$. However, for this to be a proper parameterization of $\M$, that is, for $(\epsilon_1,...,\epsilon_n)$ to define coordinates,  \eqref{Thm_parameterization_eqn3} must hold independently of the order of the maps $S^k_{\epsilon_k}$. By Frobenius Theorem,\footnote{If the integral curves of some vector fields $r_k$ ($k=1,...,n$) are coordinates, then the commutator \eqref{commutator} vanishes, since second order coordinate derivatives commute. The non-trivial part in Fobenius Theorem is the reverse implication: Given some vector fields $r_k$ which satisfy \eqref{commutator}, it follows that their integral curves define a coordinate system, c.f. \cite[Chapter 1]{Taylor} as a reference.} 
this independence of order holds if and only if the $r_k$'s commute in the sense that
\beq \label{commutator}
[r_i,r_j](\psi) \equiv r_i\big(r_j(\psi)\big) - r_j\big(r_i(\psi)\big) =0 \hspace{.5cm} \forall \psi \in C^\infty(\M).
\eeq 
Moreover, assuming there exist a coordinate system of Riemann invariants, then their defining property $r^{\,\nu}_i \partial_\nu \omega_j =0$ implies that $r^{\,\nu}_i \partial_\nu = \frac{\partial}{\partial w_i}$, for which the above Lie bracket vanishes. The reverse implication holds, since the parameters of integral curves of the $r_k$'s are Riemann invariants. 
\QED

\section{Conclusion and Outlook}

We introduced a geometric framework for the states of a conservation law lying in a Riemannian manifold and we proved that the methods of Lax and Glimm to construct solutions of the Riemann and Cauchy problem respectively can be extended to our geometric framework. For applications, this shows that it is not necessary to invert the accumulation function $g$, (which would be required to transform the conservation law to standard form), since it is possible to work in the manifolds of states. Moreover, within our framework it is possible to extend the existence result of Lax and Glimm to the case that $dg$ is \emph{not invertibe} on a finite family of co-dimension one surfaces. In this way, it is also possible to address systems of $n$ conservation laws of the form \eqref{system_cons_laws_manifold} which have \emph{fewer} than $n$ eigenvalues at co-dimension one surfaces. 

Genuine non-linearity is violated by many physical systems. We expect that one can extend Liu's construction \cite{Liu} to our geometric framework to obtain solutions of the Riemann problem when genuine non-linearity fails on co-dimension one surfaces. The uniqueness results of Bressan et al. \cite{Bressan} should apply whenever the accumulation function $g$ is invertible. Let us remark that we chose to address the Cauchy problem with Glimm's scheme over alternative methods, since Glimm's method allows for straightforward applications and provides an algorithm adapted to numerical computations. 

We believe our geometric approach to be more feasible for applying to physical conservation laws (with or without constraints), since one avoids inverting a possibly non-linear function and is only left with solving a linear generalized eigenvalue problem.  Moreover, conservation laws with an accumulation function usually arise directly from physical or chemical principles while the corresponding standard form results from a desire for notational convenience in mathematics, so that the geometric approach presented here could ultimately lead to less complicated structures which are easier to work with.  We thus believe that it is of advantage to use the geometric framework presented here for studying  conservation laws which are not readily given in standard form by physical or chemical principles.

\section*{Acknowledgements}
I am more than grateful to Prof. Dan Marchesin (IMPA) for suggesting to work on conservation laws subject to constraints of the form \eqref{constrained_system_eqn} - \eqref{constrained_system_constrained}, and for pointing out applications of these systems in multi-phase flow in porous media. I also thank him for various helpful discussions and for his enduring encouragement and guidance while I was a post-doc at IMPA. I also thank Prof. Alexei Mailybaev for pointing out reference \cite{KokubunMailybaev} to me. I thank an anonymous referee for mentioning the relation of the current paper to the relaxation systems in \cite{Bianchini,JinXin,Liu2}.

\begin{appendix} 

\section{A Short Introduction to Riemannian Geometry} \label{sec_preliminaries}

\subsection{Manifolds and Functions on Manifolds} 

A n-dimensional (topological) \emph{manifold} $\M$ is a separable Hausdorff space with the additional property that at each point $p\,\in\,M$ there exists an open neighborhood $\U \ni p$ (called a patch or coordinate neighborhood) together with a homeomorphism~$x:\,\U\rightarrow x\left(\U\right) \subset\R^n$, that is, $x$ is a continuous map such that its inverse exist on $x\left(\U\right)$ and $x^{-1}$ is continuous as well. The pair $(\U,x)$ is called a (coordinate) \emph{chart} and $x$ is often referred to as a coordinate system. The collection of all such charts (covering the manifold) is called an \emph{atlas}. If the intersection of two coordinate patches is nonempty, then $x\circ y^{-1}$ defines a mapping from an open set in $\R^n$ to another open set in $\R^n$, referred to as a coordinate transformation. If all such coordinate transformations in the atlas are $C^k$ differentiable, then the manifold $M$ is called a $C^k$-manifold and its atlas a $C^k$-atlas. 

Given two $C^k$-manifolds $\M$ and $\N$ of dimension $m$ and $n$ and a function $f:\M \rightarrow \N$, we say that $f$ is $k$-times differentiable if $y \circ f \circ x^{-1} : x(\U) \subset \R^m \rightarrow \R^n$ is a $k$-times differentiable function between $\R^m$ and $\R^n$, for all charts on $\M$, $(x,\U)$, and all charts on $\N$, $(y,\mathcal{V})$. We then write that $f \in C^k(\M,\N)$, and use the shorthand $C^k(\M)$ whenever $\N=\R$. The functions in $C^k(\M)$ are called \emph{scalars}. We call a function $f \in C^k(\M,\N)$ a $C^k$-\emph{diffeomorphism} if its inverse exists and is $k$-times differentiable as well. We say that $f \in C^k(\M,\N)$ is a \emph{local} $C^k$-diffeomorphism if for each $p \in \M$ there exists an open neighborhood $\U \ni p$ such that $f\big\rvert_\U$ is a $C^k$-diffeomorphism on its image.

\subsection{Tangent Vectors and One-Forms}

We now introduce the notion of tangent vectors at a fixed point $p\in\M$. To begin, let $\M=\R^n$ and consider a vector $v= (v^1,...,v^n)^* \in \R^n$, instead of $v$ being a $n$-tuple  we can equivalently define $v$ as the mapping $v:C^\infty(\R^n) \rightarrow \R$ for an arbitrary point $p\in\R^n$ through
\beq \label{summation_convention}
v(\psi) = \sum\limits_{\mu=1}^{n} \frac{\partial \psi}{\partial x^\mu}\bigg\rvert_p \cdot v^\mu, \ \ \ \ \ \ \ \forall \, \psi\in \, C^\infty(\R^n)  ,
\eeq
where $v^\mu$ denotes the components of $v$, for $\mu=1,...,n$. As a remark, we subsequently use the so-called \emph{Einstein summation convention} of summing over repeated upper and lower coordinate-indices, so that \eqref{summation_convention} is written as
\beq \nn
v(\psi) \equiv \ \frac{\partial \psi}{\partial x^\mu}\bigg\rvert_p \cdot v^\mu \ \ \ \ \ \ \ \forall \, \psi \in \, C^\infty(\R^n) .
\eeq
Now, it is straightforward to extend the above one-to-one correspondence between $n$-tuples and directional derivatives on $\R^n$ to manifolds. Namely, for $\psi\in C^\infty(\M)$, we define a \emph{tangent vector of $\M$ at $p$} to be a mapping 
\begin{eqnarray} \nonumber 
v:C^\infty(\M) \rightarrow \R, 
\end{eqnarray}
such that in a chart $x$ the mapping $v$ is of the form
\beq \label{tangent vector}
v(\psi) =   \frac{\partial \psi}{\partial x^\mu}\bigg\rvert_p \, (v^x)^\mu
\eeq
for some real numbers $(v^x)^\mu$, $\mu =1,...,n$, which generally depend on the coordinates.\footnote{For ease of notation, we often write that a quantity is evaluated at some $p\in\M$, even though it actually is evaluated at $x(p)\in \R^n$.} Under a change of coordinates, from $x$ to a new chart  $y$ at $p$, the chain rule gives us
\begin{eqnarray} \nonumber
v(\psi) \ = \ (v^x)^\mu \frac{\partial y^\alpha\circ x^{-1}}{\partial x^\mu}\bigg\rvert_p \,\frac{\partial \psi}{\partial y^\alpha}\bigg\rvert_p 
\ = \ (v^y)^\alpha  \,\frac{\partial \psi}{\partial y^\alpha},
\end{eqnarray}
where we define
\beq \label{tangent vector transfo}
(v^y)^\alpha \equiv  \frac{\partial y^\alpha\circ x^{-1}}{\partial x^\mu}\bigg\rvert_p  \, (v^x)^\mu.
\eeq
In light of \eqref{tangent vector transfo}, we are now able to introduce vectors again as a $n$-tuple of real numbers. Namely, given a coordinate patch $x$, a tangent vector $v$ at a point $p\in\M$ is defined as a $n$-tuple of real numbers, denoted by $(v^x)^\mu$, for $\mu=1,...,n$, such that in a coordinate patch $y$ around $p$ the tangent vector $v$ is the $n$-tuple $(v^y)^\alpha$ defined in terms of $(v^x)^\mu$ through \eqref{tangent vector transfo}. 

The fundamental relation \eqref{tangent vector transfo} is commonly referred to as a \emph{contravariant} transformation or as vectors transforming contravariantly. In this paper, we often drop the superscript $x$ and $y$ on the vector components, simply writing $v^\mu$ and $v^\alpha$ and using the type of index to indicate whether these components represent the vector in the coordinate chart $x^\mu$ or $y^\alpha$. A vector $v$ in a coordinate system $x$ is often written as 
\begin{eqnarray} \label{tangent vector basis}
v = \, v^\mu \frac{\partial}{\partial x^\mu}  \,
 \equiv \, v^\mu \partial_\mu ,
\end{eqnarray}
and the set $\partial_\mu \equiv \frac{\partial}{\partial x^\mu}$, $\mu=1,...,n$, is called the standard basis corresponding to the coordinates $x$. 

The collection of all tangent vectors at a point $p\in \M$ is called the \emph{tangent space} at $p$, which is denoted by $T_p\M$. $T_p\M$ is a $n$-dimensional vector space which is thus isomorphic to $\R^n$. In the special case $\M=\R^n$, we always identify $T_p\R^n$ with $\R^n$ through the (globally defined) standard basis on $\R^n$, which allows us to set $T_p\R^n = \R^n$ for all $p\in\R^n$. 

The above construction of tangent vectors is done at a point $p\in\M$ fixed. We define a vector-field to be a $C^k$ map that assigns to each $p\in\M$ a tangent vector in $T_p\M$, that is,
\beq \label{vectorfield_def}
v \, : p \in \M \mapsto \, v|_p \in T_p\M.
\eeq
The existence of a $C^k$ vector-field requires the manifold to be $C^k$-differentiable or higher. Given $\psi \in C^\infty(\M)$ and a vector-field $v$, then $p \mapsto v|_p(\psi)$ is a $C^k$  mapping between $\M$ and $\R$. Instead of $v|_p$ we often write $v(p)$.  

A $C^k$ curve on a manifold is a mapping $c\in C^k(I,\M)$ for some interval $I$. Given a parameterization $\epsilon \mapsto c(\epsilon)$ with $\epsilon \in I$, then the \emph{tangent vector of $c$} in some coordinate system $x^\mu$ is defined as
\beq \label{tangent_curve}
\dot{c}^\mu \equiv \frac{d x^\mu\circ c}{d\epsilon}.
\eeq
In fact, changing to different coordinates $\dot{c}$ transforms contravariantly, in the sense of \eqref{tangent vector transfo}, from which we conclude that \eqref{tangent_curve} defines a tangent vector of $\M$, that is, $\dot{c}(\epsilon) \in T_{c(\epsilon)}\M$ for each $\epsilon \in I$.   

Fixing again a point $p\in\M$, we define a \emph{one-form} at $p\in\M$ to be a linear map      
\begin{eqnarray} \nonumber
a \, : T_p\M \rightarrow \R.
\end{eqnarray}
Given coordinates $x^\mu$ at p and a vector $v= v^\mu \partial_\mu$, a one-form $a$ is determined by a $n$-tupel of real number $a_\mu$, for $\mu=1,...,n$, through
\beq \label{one-form}
a(v) = a_\mu v^\mu.
\eeq
In light of \eqref{tangent vector transfo}, for the real number defined by \eqref{one-form} to be independent of the choice of coordinates, the components $a_\mu$ must transform to new coordinates $y^\alpha$ according to  
\beq \label{one-form transfo}
(a^y)_\alpha =  \frac{\partial x^\mu\circ y^{-1}}{\partial y^\alpha}\bigg\rvert_p  \, (a^x)_\mu,
\eeq
since then
\[
(a^y)_\alpha (v^y)^\alpha =  (a^x)_\mu \,  \frac{\partial x^\mu\circ y^{-1}}{\partial y^\alpha}\bigg\rvert_p  \,     \frac{\partial y^\alpha\circ x^{-1}}{\partial x^\mu}\bigg\rvert_p  \, (v^x)^\mu = (a^x)_\mu (v^x)^\mu .
\]
We denote the collection of all one-form with $T^*_p\M$, which is again a $n$-dimensional vector space which one can thus again identify with $\R^n$.

We now introduce a notion of differentiation: Given two manifolds $\M$ and $\N$ and a function $f \in C^1(\M,\N)$, we define the \emph{differential of $f$} at $p$ as the linear map 
$$
df\big\rvert_p \, : T_p\M \rightarrow T_{f(p)}\N,
$$
which in a given coordinate chart $x^\mu$ on $\M$ and $\hat{x}^\nu$ on $\N$ maps the vector $v=v^\mu\partial_\mu \in T_p\M$ to the vector
\beq \label{differential}
df\Big\rvert_p\hspace{-0.2cm}(v) = v^\mu \, \frac{\partial (\hat{x}^\nu \circ f \circ x^{-1})}{\partial x^\mu}  \,  \frac{\partial}{\partial \hat{x}^\nu} \ \ \ \ \ \in \, T_{f(p)}\N.
\eeq
From the chain rule and \eqref{tangent vector transfo}, a straightforward computation shows that the the right hand side of \eqref{differential} is independent of the choice of coordinates.

\subsection{The Metric Tensor and Covariant Derivatives}

A \emph{metric tensor} (or simply metric) on $\M$ is a map which assigns to each $p\in\M$ a scalar product on the tangent vector space $T_p\M$, which we denote by $\langle \cdot,\cdot \rangle_\M(p)$ for each $p\in\M$. Given a coordinate system $x^\mu$ and the corresponding basis $(\partial_{\mu}|_p)_{\mu=1,...,n}$ of $T_p\M$, there exist a unique positive definite $n\times n$ matrix which represents $\langle \cdot,\cdot \rangle_\M(p)$ in the sense that for $v = v^\mu \partial_{\mu}$ and $u=u^\mu \partial_\mu$ in $T_p\M$ we have 
\beq \label{metric_representation}
\langle v,u \rangle_\M(p) = g^x_{\mu\nu}(p) v^\mu u^\nu,
\eeq
where we sum over $\mu,\nu=1,...,n$ and where $g^x_{\mu\nu}(p)$ are the real-valued ``matrix'' coefficients.\footnote{Strictly speaking, the $g^x_{\mu\nu}(p)$ are not matrix coefficient because they map a vector to a one-form, but it helps to think of them as a matrix.} In order for the representation \eqref{metric_representation} to be independent of the choice of coordinate the following transformation law between matrix coefficients $g^x_{\mu\nu}$ in coordinates $x^\mu$ and $g^y_{\alpha\beta}$ in coordinates $y^\alpha$,
\beq \label{metric_transfo}
g^x_{\mu\nu}  = g^y_{\alpha\beta} \frac{\partial y^\alpha\circ x^{-1}}{\partial x^\mu} \frac{\partial y^\beta\circ x^{-1}}{\partial x^\nu},
\eeq
where we sum over $\alpha,\,\beta=1,...,n$. It follows from \eqref{tangent vector transfo} and \eqref{metric_transfo} that the right hand side of \eqref{metric_representation} is independent of the choice of coordinate. We say that a metric tensor is in $C^k$ if $g^x_{\mu\nu} \in C^k(\M,\R)$ for all coordinate systems in the atlas. A manifold endowed with a metric tensor is called a \emph{Riemannian manifold}, which is the subject of Riemannian geometry.

We now introduce \emph{covariant derivatives}, which gives us the important notion of derivatives of vector fields and tensors. To begin, consider some vector field in a coordinate system, $v=(v^x)^\mu \partial_{x^\mu}$. Naively taking a partial derivative of $v$ with respect to $x^\nu$ in the direction of another vector field $w$ would result in an expression which is no tangent vector, since it does not transform according to \eqref{tangent vector transfo} but yields additional second order derivative terms of the coordinate transformation. The remedy to this problem is to augment partial derivatives with terms which cause cancellation under coordinate transformations. These terms are the so-called \emph{Christoffel Symbols}, defined in terms of the metric tensor as
\beq \label{Christoffel}
\Gamma^\mu_{\sigma\rho} \equiv g^{\mu\nu} \left( \frac{\partial g_{\nu\rho}}{\partial x^\sigma} + \frac{\partial g_{\nu\sigma}}{\partial x^\rho} - \frac{\partial g_{\sigma\rho}}{\partial x^\nu}  \right) = \Gamma^\mu_{\rho\sigma}.
\eeq
The covariant derivative $\nabla$ of $v$ in the direction of $w$ in coordinates $x^\mu$ is now defined as
\beq \label{covariant_derivative}
\nabla_w v^\mu 
\equiv \frac{\partial v^\mu}{\partial x^\nu} w^\nu + \Gamma^\mu_{\sigma\rho} w^\sigma v^\rho.
\eeq
A straight forward computation, using the transformation laws \eqref{tangent vector transfo} and \eqref{metric_transfo} together with \eqref{Christoffel}, shows that $\nabla_w v^\mu$ transforms by \eqref{tangent vector transfo}, from which we conclude that the definition \eqref{covariant_derivative} indeed defines a vector $\nabla_w v|_p \in T_p\M$. The covariant derivative generalizes partial derivatives and has the following properties: Let $u,\,w$ be vector fields and $f\in C^1(\M)$, then  
\begin{eqnarray}
\nabla_{w+fu}v &=& \nabla_w v + f\nabla_u v \label{covariant_derivtive_linearity1} \\
\nabla_{w}(v+u) &=& \nabla_w v + \nabla_w u \label{covariant_derivtive_semi-linearity} \\
\nabla_w(fv) &=& w(f) v + f \nabla_w v \label{covariant_derivtive_Leipnitz-rule} \\
w\left( \langle u,v\rangle_\M \right) &=& \langle \nabla_w u, v\rangle_\M + \langle u, \nabla_w v\rangle_\M . \label{covariant_derivtive_metric} 
\end{eqnarray}
$\nabla$ is also called the \emph{Levi-Civita-connection}. IN general, a connection on $\M$ refers to a mapping from $T_p\M$ into $T_p\M$ which satisfies \eqref{covariant_derivtive_linearity1} - \eqref{covariant_derivtive_Leipnitz-rule}.  Various connections can be defined on $\M$, but the Levi-Civita-connection is the unique connection which satisfies \eqref{covariant_derivtive_metric} and therefore plays a key-role in Riemannian geometry. 

As a slight variation of the above, we define covariant differentiation of a vector field $v$ along some curve  $c: \epsilon \mapsto c(\epsilon) \in \M$ as
\beq \label{covariant_derivative_curve}
 \nabla_{\frac{\partial}{\partial \epsilon}} v^\mu  \
 \equiv \ \frac{d v^\mu}{d \epsilon} + \Gamma^\mu_{\sigma\rho}\, \dot{c}^\sigma\, v^\rho
 \ = \ \frac{\partial v^\mu}{\partial x^\sigma}\, \dot{c}^\sigma + \Gamma^\mu_{\sigma\rho}\, \dot{c}^\sigma\, v^\rho
\eeq
where $v^\mu$ and $\Gamma^\mu_{\sigma\rho}$ are restricted to $c$. Moreover, we define the covariant derivative of a one-form $a_\mu$ in the direction of a vector field $v$ as
\beq \label{covariant_derivative_oneforms}
\nabla_v a_\mu 
\equiv \frac{\partial a_\mu}{\partial x^\nu} v^\nu - \Gamma^\sigma_{\mu\rho} v^\rho a_\sigma,
\eeq
for which the analogous rules in \eqref{covariant_derivtive_linearity1} - \eqref{covariant_derivtive_metric} hold again. Finally, the covariant derivative of a function $f$ is defined as $\nabla_v f \equiv v(f)$.

\subsection{Geodesic Curves and Arc-Length Parameterization}

A \emph{geodesic curve} is a differentiable curve $c: (a,b)\rightarrow \M$ which tangent vector has a vanishing covariant derivative along itself, $\nabla_{\frac{\partial}{\partial \epsilon}} \dot{c} = 0$. That is, expressed in coordinates $x^\mu$ a geodesic curve solves the ODE
\beq \nn
\frac{d^2 c^\mu}{d\epsilon^2} + \Gamma^\mu_{\sigma\rho} \frac{dc^\sigma}{d\epsilon} \frac{d c^\rho}{d\epsilon} = 0.
\eeq
From \eqref{covariant_derivtive_metric} and $\nabla_{\frac{\partial}{\partial \epsilon}} \dot{c} = 0$ it follows that $\langle \dot{c},\dot{c} \rangle_\M(\epsilon)$ is constant. Geodesic curves are the shortest curves connecting two points as long that they are sufficiently close together. Thus, geodesics locally minimize the canonical distance $d_\M$ defined in \eqref{distancefunction}.

Given a curve $\epsilon\mapsto c(\epsilon) \in \M$, we say it is \emph{parameterized by arc-length} if $\langle\dot{c},\dot{c}\rangle_\M(\epsilon)=1$ for all $\epsilon$. In this parameterization it follows that $\int_a^b |\dot{c}|_\M(\epsilon) d\epsilon = b-a$. Any $C^1$ curve with non-vanishing tangential derivative can be parameterized by arc-length, namely, given a parameterization $\epsilon'\mapsto c(\epsilon')$, define the arc-length parameter $\epsilon(\epsilon')$ by solving the ODE 
\beq \label{arc-length_change}
\frac{d\epsilon}{d\epsilon'} = \sqrt{\left|\frac{dc}{d\epsilon'}\right|_\M}
\eeq
which implies 
\beq \label{arc-length_verification}
\left|\frac{d c}{d\epsilon}\right|_\M = \left|\frac{d c}{d\epsilon'}\right|_\M  \left|\frac{d\epsilon'}{d\epsilon}\right|^2 = 1.
\eeq
Arc-length parameterization implies that the curve does not undergo any acceleration along itself, as can be seen from the following computation. Differentiating the normalization $\langle\dot{c},\dot{c}\rangle_\M=1$ and using the Leipnitz rule of the Levi-Civita connection, we find
\beq \label{arc-length_acceleration}
0 = \frac{d}{d\epsilon} \langle\dot{c},\dot{c}\rangle_\M = 2 \langle\dot{c},\ddot{c}\rangle_\M,
\eeq
which shows the acceleration of the curve being orthogonal to its velocity.

\subsection{Submanifolds}

A $n$-dimensional \emph{submanifold} $\M$ of $\R^m$, $m\geq n$, is subset of $\R^m$ which is a $n$-dimensional manifold such that the topology of $\M$ is the induced topology of $\R^m$, that is, $\U\subset\M$ is an open set if and only if there exists an open set $U$ in $\R^m$ such that $\U = U \cap \M$. 

Given a differentiable function $C:\R^m \rightarrow \R^n$, for $n<m$, such that $DC|_p$ has full rank whenever $C(p)=0$, then the set $C^{-1}(0)$ defines a submanifold of $\R^m$, \cite[Theorem 5.1]{Spivak}. In more detail, assuming that $p\in \hat{\U}$ with $C(p)=0$, then, since $C$ is assumed to have full rank at $p$, the implicit function theorem yields that there exists an open set $\Omega \subset \R^n$ and a differentiable function $\varphi_{p}: \Omega \rightarrow \R^m$ invertible on its image such that $\varphi_{p}$ parameterizes the solution space $C^{-1}(0)$ sufficiently close to $p$, c.f. \cite[Theorem 2.11]{Spivak}. Since the implicit function theorem yields for any point $p \in \hat{\U}$ with $C(p)=0$ such a function $\varphi_{p}$ parameterizing some open subset $\U_p$ in $C^{-1}(0)$, we obtain an open covering of $C^{-1}(0)$.  We conclude that 
\beq \nn
\M \equiv \big\{p\in \R^m:\, C(p)=0\big\}
\eeq 
is a $n$-dimensional submanifold of $\R^m$. The tangent space of $\M$ is given by 
\beq \nn
T_p\M = \text{nul} \left(DC|_p\right),
\eeq
since the derivative of any curve $\gamma$ lying in $\M$ and passing through $p$ satisfies 
\beq\nn
0 = \frac{d}{d\epsilon}C(\gamma(\epsilon)) = DC\Big|_{\gamma(\epsilon)} \left(\frac{d\gamma}{d\epsilon}\right)
\eeq
and for each $v\in T_p\M$ exist a curves through $p$ which tangent vector at $p$ is identical to $v$.

\end{appendix}

\providecommand{\bysame}{\leavevmode\hbox to3em{\hrulefill}\thinspace}
\providecommand{\MR}{\relax\ifhmode\unskip\space\fi MR }
\providecommand{\MRhref}[2]{%
  \href{http://www.ams.org/mathscinet-getitem?mr=#1}{#2} }
\providecommand{\href}[2]{#2}

\end{document}